\documentclass[twocolumn]{elsart}

\usepackage{graphicx}
\usepackage{amssymb,amsmath}

\begin{document}

\begin{frontmatter}

\journal{Nucl. Instr. and Meth. A}

\title{Development of an atmospheric Cherenkov imaging camera for 
the CANGAROO-III experiment
}

\author[1]{S.~Kabuki},
\author[1]{K.~Tsuchiya},
\author[1]{K.~Okumura},
\author[1]{R.~Enomoto\corauthref{cor1}},
\author[1]{T.~Uchida},
\author[1]{H.~Tsunoo},
\author[2]{Shin.~Hayashi}, 
\author[2]{Sei.~Hayashi}, 
\author[2]{F.~Kajino}, 
\author[2]{A.~Maeshiro}, 
\author[2]{I.~Tada},
\author[3]{C.~Itoh},
\author[4]{A.~Asahara},
\author[5]{G.V.~Bicknell},
\author[6]{R.W.~Clay},
\author[7]{P.G.~Edwards},
\author[8]{S.~Gunji},
\author[4,12]{S.~Hara},
\author[9]{T.~Hara},
\author[10]{T.~Hattori},
\author[1]{H.~Katagiri},
\author[1]{A.~Kawachi},
\author[11]{T.~Kifune},
\author[4]{H.~Kubo},
\author[4,12]{J.~Kushida},
\author[13]{Y.~Matsubara},
\author[14]{Y.~Mizumoto},
\author[1]{M.~Mori},
\author[10]{H.~Moro},
\author[15]{H.~Muraishi},
\author[13]{Y.~Muraki},
\author[9]{T.~Naito},
\author[10]{T.~Nakase},
\author[4]{D.~Nishida},
\author[10]{K.~Nishijima},
\author[1]{M.~Ohishi},
\author[6]{J.R.~Patterson},
\author[6]{R.J.~Protheroe},
\author[12]{K.~Sakurazawa},
\author[6]{D.L.~Swaby},
\author[4]{T.~Tanimori},
\author[8]{F.~Tokanai},
\author[8]{A.~Watanabe},
\author[4]{S.~Watanabe},
\author[3]{S.~Yanagita},
\author[3]{T.~Yoshida},
\author[16]{T.~Yoshikoshi}

\address[1]{Institute for Cosmic Ray Research, University of Tokyo,
 Chiba 277-8582, Japan}
\address[2]{Department of Physics, Konan University, 
Hyogo 658-8501, Japan}
\address[3]{Faculty of Science, Ibaraki University, Ibaraki
 310-8512, Japan}
\address[4]{Department of Physics, Kyoto University, Sakyo-ku, Kyoto 606-8502, Japan}
\address[5]{MSSSO, Australian National University, ACT 2611, Australia}
\address[6]{Department of Physics and Math. Physics, University of Adelaide, SA 5005, Australia}
\address[7]{Institute of Space and Astronautical Science, Sagamihara, Kanagawa 229-8510, Japan}
\address[8]{Department of Physics, Yamagata University, Yamagata, Yamagata 990-8560, Japan}
\address[9]{Faculty of Management Information, Yamanashi Gakuin University, Kofu,Yamanashi 400-8575,
 Japan}
\address[10]{Department of Physics, Tokai University, Hiratsuka, Kanagawa 259-1292, Japan}
\address[11]{Faculty of Engineering, Shinshu University, Nagano, Nagano 380-8553, Japan}
\address[12]{Department of Physics, Tokyo Institute of Technology, Meguro-ku, Tokyo 152-8551, Japan}
\address[13]{STE Laboratory, Nagoya University, Nagoya, Aichi 464-8601, Japan}
\address[14]{National Astronomical Observatory of Japan, Mitaka, Tokyo 181-8588, Japan}
\address[15]{Department of Radiological Sciences, 
Ibaraki Prefectural University of Health Sciences,
Ibaraki 300-0394, Japan
}
\address[16]{Department of Physics, Osaka City University, Osaka, Osaka 558-8585, Japan}

\begin{abstract}

A Cherenkov imaging camera for the CANGAROO-III experiment 
has been developed
for observations of gamma-ray induced air-showers 
at energies from 10$^{11}$ to 10$^{14}$~eV.
The camera consists of 427 pixels, arranged in a hexagonal shape 
at 0.17$^\circ$ intervals, each of which is a 
3/4-inch diameter photomultiplier module
with a Winston-cone--shaped light guide.
The camera was designed to have a large dynamic range of 
signal linearity, a wider field of view, and
an improvement in photon collection efficiency compared with the
CANGAROO-II camera.
The camera, and a number of the calibration experiments made to
test its performance, are described in detail in this paper.

\end{abstract}


\corauth[cor1]{
Corresponding author. 
{\it E-mail address:} enomoto@icrr.u-tokyo.ac.jp,
ICRR, Univ. of Tokyo, 5-1-5 Kashiwano-ha, Kashiwa, Chiba 277-8582,
Japan, tel: +81-4-7136-5116, fax: +81-4-7136-3133.
}

\end{frontmatter}


\newpage

\section{Introduction}

CANGAROO (the Collaboration of Australia and Nippon for a GAmma Ray
Observatory in the Outback)
is a project to study TeV gamma-ray emitting celestial sources,
particularly those in the southern sky.
The earlier experiments CANGAROO-I and CANGAROO-II, 
using single
telescopes of 3.8-meter~\cite{hara93} and 10-meter diameter, 
respectively,
obtained glimpses of astronomical objects 
such as pulsars (PSR~1706-44~\cite{kifune95}, Crab~\cite{tanimori98a}),
active galactic nuclei (Mrk~421~\cite{okumura02})
and supernova remnants (SN1006~\cite{tanimori98b}, 
RXJ1713.7-3946~\cite{muraishi00,enomoto02b}).

CANGAROO-III, the next phase of the project, aims at the
detection at energies of 0.1$\sim$100~TeV (1TeV=10$^{12}$eV) 
using four 10-meter-diameter telescopes for 
stereoscopic (and ``multi-scopic'')
reconstruction of atmospheric Cherenkov shower images.
Various improvements in the design of the imaging Cherenkov camera 
have been made for the CANGAROO-III 
experiment~\cite{morim01,kubo01,kajino01}.

The first telescope of the CANGAROO-III array is the
CANGAROO-II telescope, which has been in operation since April, 2000. 
The telescope has an optically parabolic shape, and
consists of 114 small spherical mirrors with diameters of 80~cm and 
made of carbon-fibre reinforced plastic~\cite{kawachi01}. 
The second telescope was built in March 2002 and
the remaining two telescopes will be constructed by the end of the
2004.

The technique of stereoscopic observing~\cite{hegra97,enomoto02a}, 
allows more precise measurements to be made:
a significant improvement in the separation efficiency 
of gamma-ray events from the cosmic-ray background,
a more precise reconstruction of the arrival direction 
determined on an event-by-event basis~\cite{akerlof91}, and
an improvement in the energy resolution of up to 20\%~\cite{hoffman97}.
The stereoscopic technique is effective for both point-like sources
and for extended gamma-ray sources, such as supernova remnants.

Measurement of the gamma-ray energy spectrum over a wide range 
is important for determining the gamma-ray emission mechanism.
Information derived from  spectral features,
such as the spectral index and any cutoff energy, can help clarify 
the acceleration and production mechanisms,
and also provide estimations of fundamental parameters, 
such as the magnetic field and the maximum energy of the
particles accelerated at the source~\cite{tanimori98b}.
For instance, for the Crab nebula, 
the detection of gamma-rays above 50~TeV is a key 
to finding the acceleration limit
in the pulsar nebula~\cite{tanimori98a}.
Also, for supernova remnants, measurement of the energy spectrum 
at sub-TeV energies has provided
evidence for proton acceleration~\cite{enomoto02b}.
Spectral measurements are crucial 
for solving the mystery of the origin of cosmic rays.

Another goal of the experiment is to extend observations
into the energy region between $\sim$10~GeV 
(below which was explored by the EGRET detector on 
the Compton Gamma-Ray Observatory~\cite{EGRET})
and the typical Cherenkov telescope threshold to date of 200--300~GeV. 
A number of the astronomical objects observed by EGRET
are expected to be detectable with Cherenkov telescopes
by narrowing this gap.
A lowering of the energy threshold will also allow 
the detection of new gamma-ray sources, 
especially those with steep energy spectra,
such as distant active galactic nuclei, for which
sub-TeV gamma-rays are less attenuated by 
the infrared background than those in the 
TeV region~\cite{hikishov62,gould67,stecker92}.

In this paper, the development of the Cherenkov imaging camera
for the second CANGAROO-III telescope is described.
Detailed studies were made concerning 
the number, size, type and arrangement of photomultiplier tubes, 
the preamplifier circuits, the light guides,
reliable calibration sources and 
the overall performance of the camera system.


\section{Miscellaneous developments}

The weight of the camera was constrained 
by the mechanical structure of the telescope
to be less than $\sim$100~kg.
Based on a structural analysis,
a deformation of  $\lesssim$2~mm at the focal plane would be
expected for a 100~kg camera.

Images of gamma-ray showers are located in the focal plane
at distance of up to $\sim1.5^\circ$ from the source position.
For stereoscopic observations, 
some refined observation modes have been proposed, such as 
a  ``raster scan'' tracking mode, with which the telescope
tracking centers simultaneously scan a square region of $\pm$0.5$^{\circ}$ in
right ascension and declination coordinate centered on the target
\cite{hayashida98}, 
a ``wobble'' mode in which the telescopes are pointed
at a direction displaced from the source by first $\sim0.5^\circ$
and then $\sim -0.5^\circ$~\cite{lampeitl99}, 
and a ``convergent'' mode~\cite{grindlay75}.
The field of view of the CANGAROO-II telescope was
$\sim$2.7$^\circ$, however from a consideration of the points above
we have chosen a wider field of view, $\sim4^\circ$,
for the remaining CANGAROO-III telescopes. 

Differentiation of gamma-ray and cosmic-ray showers
is based on the shape and orientation parameters of the images, such
as $length$, $width$, $distance$ 
and $alpha$~\cite{hillas82,reynolds93}.
These are calculated from the amplitudes of the signals detected
by the pixels in the focal plane.
In order to accurately reconstruct the shower image,
good imaging performance, with negligible cross-talk between the pixels
and uniform gains for all of the pixels, is required.
Also, as a camera with a wider field of view
has an increased chance of having a bright star within the field of view,
the high voltages (HV) 
of the camera pixels should be individually controllable,
to avoid possible 
deterioration of photomultiplier tube performance by
the higher trigger rates that bright starts cause.

Gamma-ray cascades,
which develop in the upper atmosphere, 
produce light pools with radii of about 120~meters on the ground 
with a temporal spread of about 10~ns.
A high-sensitivity photomultiplier tube (PMT), 
having a gain (after pre-amplification) of $\sim2\times10^7$, and some 
sensitivity to ultraviolet photons,
is a suitable detector of Cherenkov photons.
It should also be robust and stable, particularly with regard
to the night sky background (NSB),
since a significant, and time-variable, number of photons 
are received from the night sky, starlight, and terrestrial sources. 
A wide dynamic-range, from 1 to 250~photoelectrons (p.e.) with less than
10\% deviation from linearity, for each sensor
is required for precise measurements of energy spectra.
Also, maximizing the collection efficiency of
Cherenkov photons is required to reduce the energy threshold.

A timing resolution of less than 1~ns is important for 
discriminating Cherenkov photons of air showers 
from the NSB, 
and is also crucial for the event discrimination of 
gamma-rays from hadrons 
using relative arrival time information~\cite{muraishi00}.

\section{Camera design}

\subsection{General design}

\begin{figure}
\begin{center}
\begin{minipage}{50mm}
\includegraphics[width=50mm]{camera-plane-design.epsi}
\end{minipage}
\begin{minipage}{50mm}
\includegraphics[width=50mm]{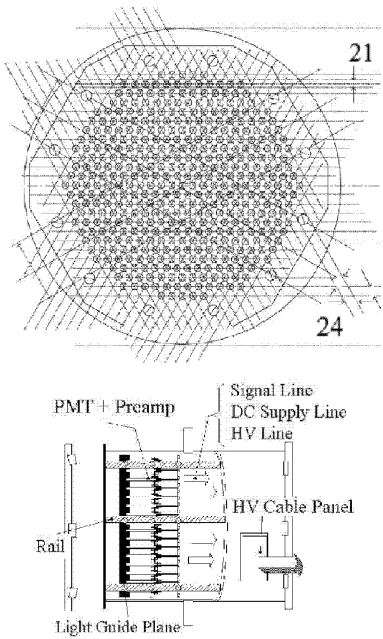}
\end{minipage}
\vskip 1cm
\caption{Schematic design of the CANGAROO-III camera frame.
The upper figure shows the front view and
the lower figure shows a side view of the camera.
The camera pixels are arranged on a hexagonal grid,
as shown in the upper figure.}
\label{fig:cameraplane}
\end{center}
\end{figure}

The camera design of 
the second CANGAROO-III telescope was made
based on the above requirements.
A schematic diagram of the camera structure 
is shown in Fig.~\ref{fig:cameraplane}.

The camera is contained in a cylindrical vessel
of 800~mm in diameter and 1000~mm in length,
which provides shielding from both rain and light.
The vessel is made of an aluminum alloy (A5052)
in order to reduce the weight and provide sufficient rigidity.
Inside the camera vessel, 427 PMT modules,
regulator circuit panels,
an LED (light emitting diode) light diffuser for gain calibration,
and several other instruments such as a thermometer, 
are contained.
Twisted-pair cables for transmitting signals 
and multi-wired cables for the high-voltage supply are
fed into the camera vessel.

The camera frame consists of two  aluminum 
templates
(5~mm in thickness),
in which holes (21~mm in diameter) are drilled
at locations corresponding to each of the 427 pixels, 
as shown in Fig.~\ref{fig:cameraplane} (upper).
Every PMT module, consisting of an PMT and a preamplifier
(20.5~mm in diameter), is held by these templates.
Light guides are attached to the front panel.
The photocathode plane of the PMT module is held close to
the back plane of the light guide.
The front panel, on which the light guides are attached, 
is fixed at the focal plane.
All segmented mirrors can be seen from every pixel position.

The pixels are arranged in hexagonal shape in order to maximize the
collection efficiency of the Cherenkov light.
The pixel size was determined to be 0.17 degrees from a simulation 
study~\cite{enomoto02a}, 
taking into account 
the spot size of the 10\,m composite mirror~\cite{kawachi01}.
The simulation code was based on GEANT 3.21. It included the electro-magnetic
shower and the telescope optics.


\subsection{Photomultiplier module}

\begin{figure}
\begin{center}
\includegraphics[width=50mm]{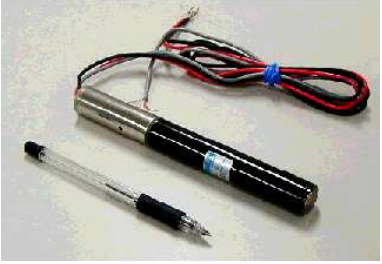}
\vskip 1cm
\caption{Photograph of the PMT module. }
\label{fig:pmtmodule}
\end{center}
\end{figure}

A photograph of the PMT module is shown in Fig.~\ref{fig:pmtmodule}.

It is cylindrical with a 
diameter of 20.5~mm and length of 173.5~mm.
Three types of cable (the signal, the D.C. power and the high voltage supply)
are passed through from the back end of the module.
The module consists of a 19\,mm (3/4\,inch) PMT (Hamamatsu R3479UV), 
bleeder circuits and a pre-amplifier which are attached at the base of
the PMT.
An operational amplifier (MAX4107) is used as the pre-amplifier and
the input signal from the PMT is amplified by a factor of 60.
The bleeder circuit is coated with epoxy resin in order to prevent
humidity-related effects.
Each module is shielded with $\mu$-metal (0.2~mm thick)
to reduce any effect due to the geomagnetic field and 
cross-talk between PMT modules.
Each module (without cables) weighs 75~g.

\subsubsection{Photomultiplier tube selection}

The 19\,mm diameter PMT (15~mm diameter photocathode) 
was chosen considering the field of view the camera.
It meets the requirements of:

\begin{itemize}
\item a high sensitivity for light at ultraviolet wavelengths,
\item a gain (before pre-amplification) of greater than 10$^5$,
\item a transit time spread (T.T.S.) of less than 1~ns,
\item a wide range of the linear response for 1$\sim$300~photoelectrons
      with less than 20\% deviation from linearity,
\item a good signal/noise separation and possible identification of 
      the single photoelectron peak, and
\item a light weight and compact size.
\end{itemize}

\begin{figure}
\begin{center}
\includegraphics[width=40mm,angle=270]{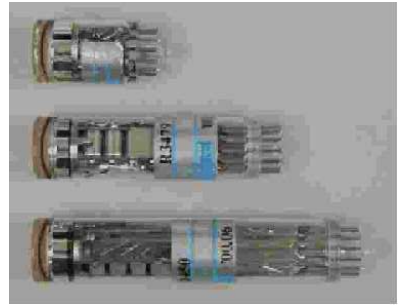}
\vskip 1cm
\caption{Performance tests were made for three
types of 19mm (3/4 inch) diameter photomultiplier tubes (PMTs).
The model numbers are, from top to bottom,
Hamamatsu Photonics R5611, R3479 and R1450, respectively.
The R3479 was selected for the CANGAROO-III PMT module.}
\label{fig:photomul}
\end{center}
\end{figure}

Three types of 19\,mm PMT (Hamamatsu R1450, R3479 and R5611), 
shown in Fig.~\ref{fig:photomul},
were compared for selection.

The basic characteristics of these PMTs are 
listed in Table~\ref{table:pmtlist}.

\begin{table}[b]
\begin{tabular*}{70mm}{p{20mm}lrrr}
\hline
\hline
 Model no.                       &  R1450  &  R3479 &  R5611 \\
\hline
Dynode structure                &    LINE &   LINE &     CC \\
No. of dynode stages            &      10 &      8 &      8 \\
Dynamic range~(p.e.)            &  $>$150 & $>$150 &     50 \\
Weight~(g)                      &      19 &     15 &      9 \\
Length~(cm)                     &     8.8 &    6.5 &    3.0 \\
Supplied high voltage ($8\times10^5$~gain)   &  1170 & 1500  & 780 \\
Rise time~(nsec)                &     1.8 &    1.3 &    1.5 \\
Transit time~(nsec)             &      19 &     14 &     17 \\
Transit time spread FWHM~(nsec) &    0.76 &   0.36 &    0.8 \\
\hline
\hline
\end{tabular*}
\vskip 1cm
\caption{Comparison of the performances of three types of PMTs.
LINE means linear-focused type of dynode and CC means circular-cage type.}
\label{table:pmtlist}
\end{table}

The R5611 has a circular-cage dynode, 
which is superior in compactness and lower voltage operation,
whereas the R3479 and R1450 have linear-focused dynodes.
All are light weight, with fast T.T.S.s, and 
allow the identification of a single photoelectron peak.
The choice of PMT was ultimately 
based on the linear response over a wide dynamic range.

Linearity was checked using a blue LED 
(NSPB510S, Nichia, $\lambda\sim470$-nm)~\cite{khan98}
which
was flashed with a 20~ns wide pulse 
from a pulse generator (AVI-V-C-N, AVTECH).
The high-voltage supplied to each PMT 
was adjusted to give a gain of $2\times10^5$,
which was determined by measurement of the single p.e. peak.

The amount of light was changed in the range of $1\sim250$~p.e.
using combinations of neutral density (ND) filters
(MAN-52, SIGMA Koki Optical Instruments) with different transmission
efficiencies.
The optical attenuation of each filter was checked to be within $\pm$1\% 
accuracy.
The signal outputs were amplified by 10 
and then digitized by a charge-integrated ADC
(Analog-Digital-Converter) (0.24~pC/ch) with a 100-ns gate width.
Here, we used a commercial amplifier of PHILLIPS (model 776).
Its linearity was checked in advance.
Trigger signals were made using the output signals of the pulse generator.

\begin{figure}
\begin{center}
\begin{minipage}{50mm}
\includegraphics[width=50mm]{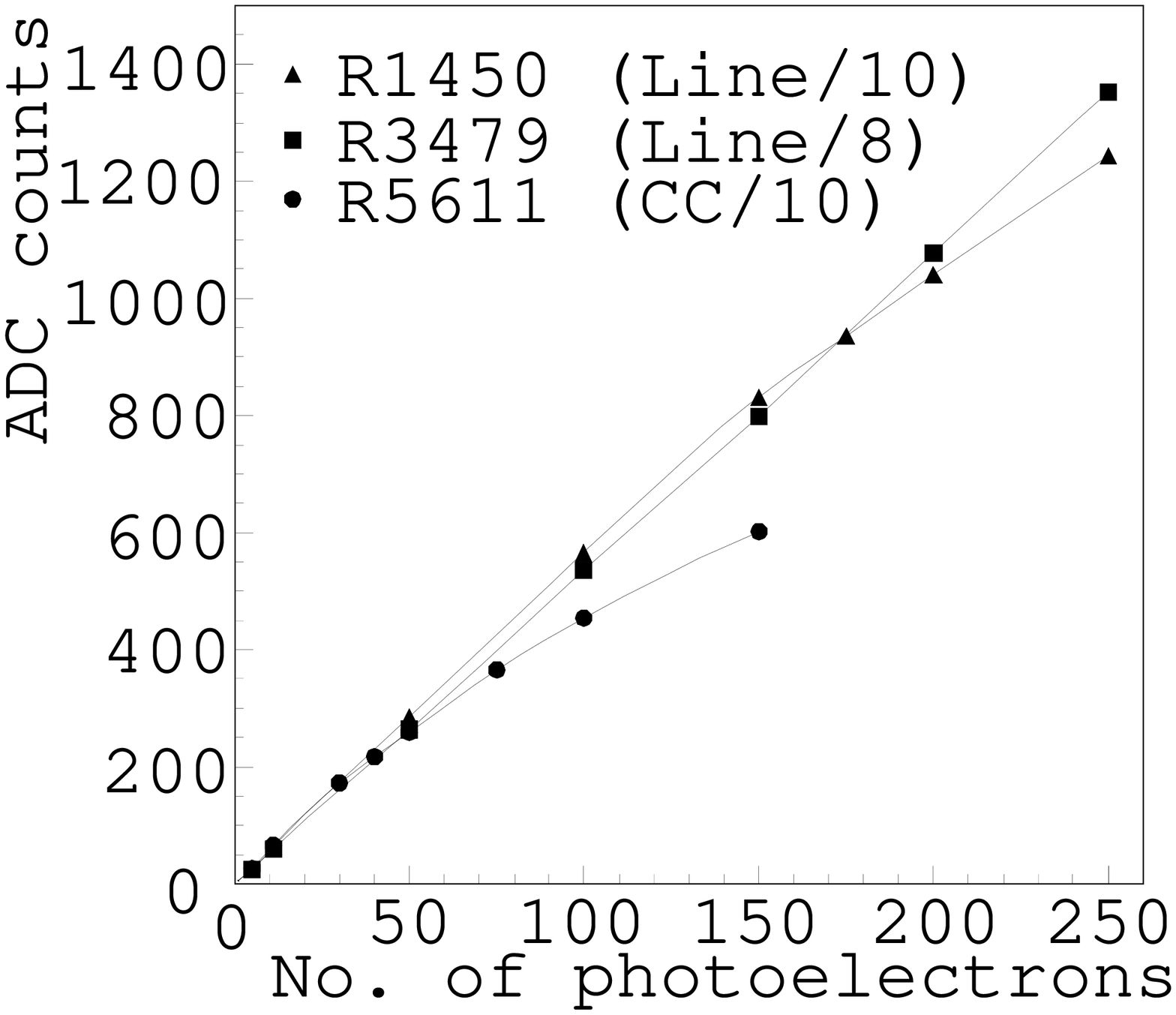}
\end{minipage}
\begin{minipage}{50mm}
\includegraphics[width=50mm]{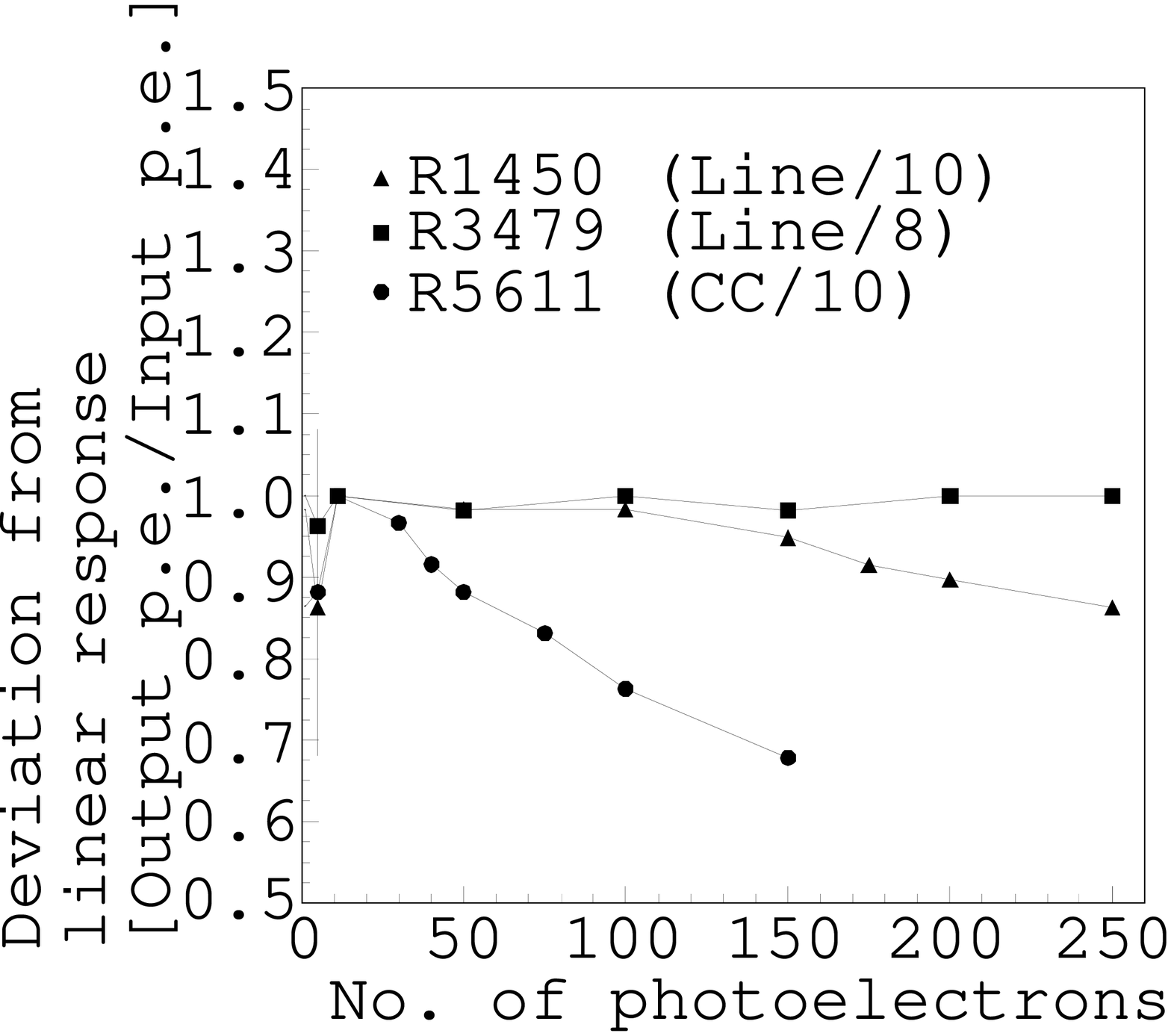}
\end{minipage}
\vskip 1cm
\caption{ADC counts (upper) and deviation from linear response (lower)
as a function of number of input photoelectrons.
In the lower figure,
the large errors at a small input p.e.
are considered to be due to the uncertainty in  the input
number of photoelectrons.  }
\label{fig:PMTT_linearity_comparison}
\end{center}
\end{figure}

The measured linearities, ADC counts and deviations from linear
response as a function of the number of input photons,
are shown in Fig.~\ref{fig:PMTT_linearity_comparison}.

The circular-cage type (R5611) was found to have
less linearity at a higher levels of incident light.
This is due to the characteristics of the wire structure of 
the anode (space-charge limitation). 
Meanwhile, the linearity of the R3479 shows a slightly better result than 
that of the R1450.
From this result, and 
also the considerations of compactness and weight,
the R3479 was determined to be the most suitable for our experiment, 

UV glass, which has a better transparency for ultraviolet
wavelengths, is used for the photocathode window.
The quantum efficiency is discussed further in
section~\ref{section:quamtum_efficiency}.

\subsubsection{Preamplifier circuit}
\label{section:preamp}

The characteristics of a fast rise-time and 
a high band-width are required for the preamplifier
for signal amplification.
Before designing the preamplifier circuit,
three types of the preamplifier chips,
CS510S (Clear Pulse)~\cite{tsukada91}, $\mu$pc1664c (NEC) 
and MAX4107~(MAXIM), 
were tested, to study in particular the linearity of the gain.

\begin{figure}
\begin{center}
\includegraphics[width=50mm]{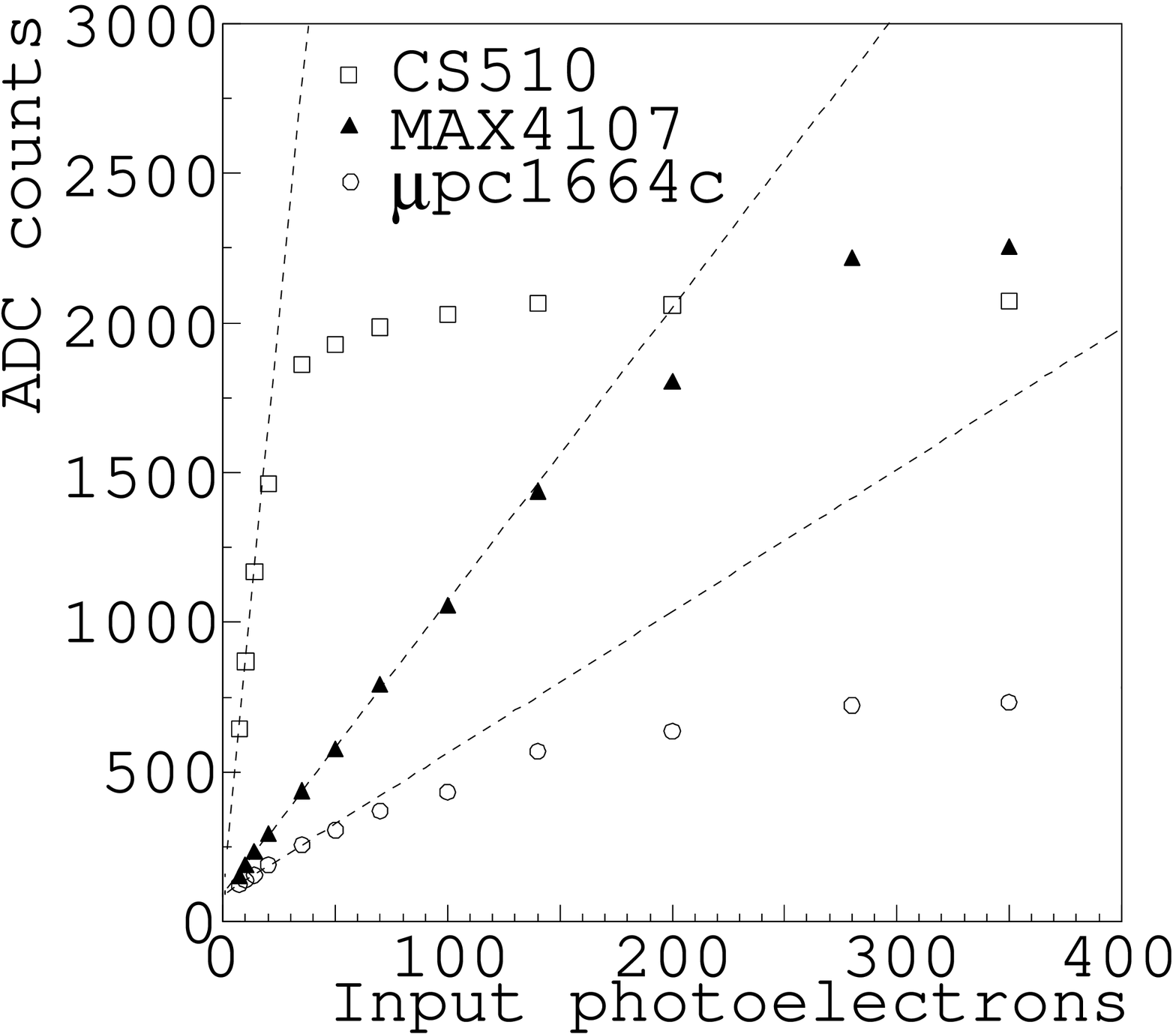}
\vskip 1cm
\caption{Linearities for three types of preamplifier chips
(CS510S, $\mu$pc1664c and MAX4107).}
\label{fig:preamp-dynamicrange}
\end{center}
\end{figure}

The results of the linearity measurement are shown in 
Fig.~\ref{fig:preamp-dynamicrange}.

Here, we used R3479 for input of the preamp without the $\times$10-amplifier.
The amount of input light was adjusted by combinations of ND filters,
as in the previous measurement.
From Fig~\ref{fig:preamp-dynamicrange}, 
it is found that MAX4107 has the widest dynamic range 
of the three tested, and so it was selected for the photomultiplier module.

\begin{figure}
\begin{center}
\includegraphics[width=70mm]{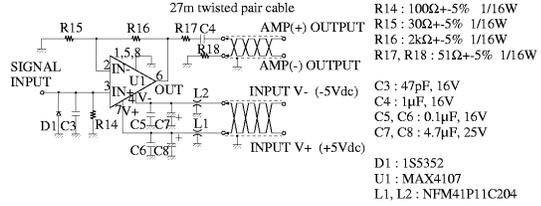}
\vskip 1cm
\caption{Schematic diagram of the preamplifier circuit.}
\label{fig:preamp-circiut}
\end{center}
\end{figure}

A schematic diagram of the preamplifier circuit is shown 
in Fig.~\ref{fig:preamp-circiut}.

The feedback resistances are selected by adjusting the 
proper 
gain of the preamplifier.
The rise time of the pulse signal is smeared to $\sim$5~ns
at the signal input (C3 in Fig.~\ref{fig:preamp-circiut})
to simulate 
transmission in the signal cables.
As the duration of the
Cherenkov light from air showers is of the same order,
this smearing has the advantage of reducing the peak voltage of 
the signal, allowing a wider dynamic-range of the linearity
to be obtained.

\subsubsection{High-voltage polarity}

The positive polarity of the high voltage is 
supplied to the PMTs (cathode-ground).
The advantage of a cathode-ground is that it
prevents discharges between the cathode plane of the PMT and
the metal structure of the camera.
The light-guide can be attached directly at the photocathode plane 
of the PMT module.

In the CANGAROO-II camera~\cite{morim99},
a polycarbonate light-guide was attached to the PMT window where
the negative polarity is applied (anode-ground).
Frequent discharges were observed on nights of high humidity (especially in
winter).
Some space ($\sim$2~mm) between the photocathode was required,
which significantly decreased the collection efficiency. 
Also, this HV configuration was found to be the cause 
of an increase in the noise rate
at the single-photon level
because of an instability of the electric potential, as shown
in the following measurement.

\begin{figure}
\begin{center}
\includegraphics[width=50mm]{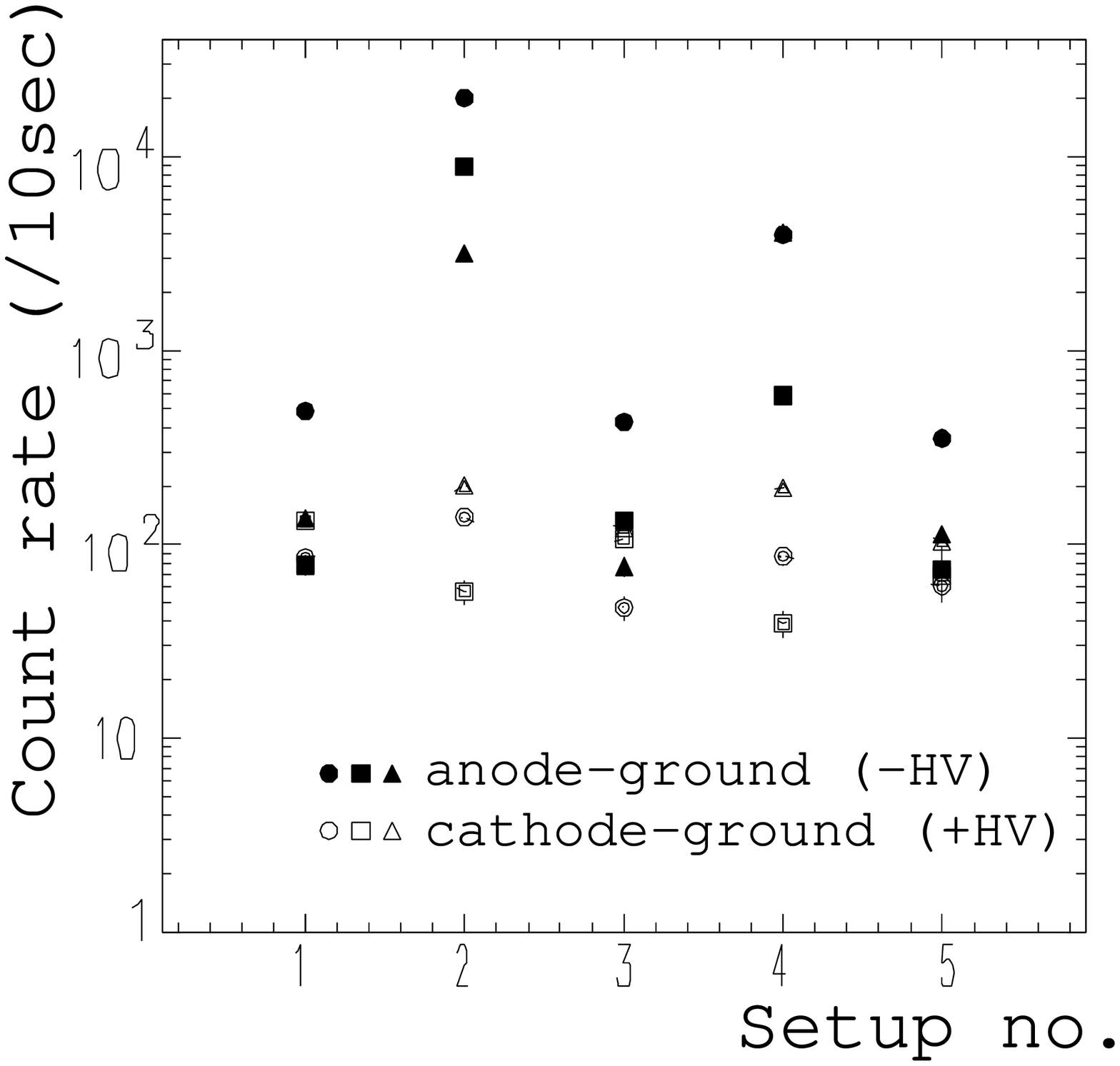}
\vskip 1cm
\caption{Count rates for different PMT setups.
The filled marks (circles, squares and triangles) show the results with
the anode-ground setup (negative polarity of the supplied voltage),
and the open marks show that of the cathode ground (positive polarity).
See text for details of the setups.}
\label{fig:noiserate_discharge}
\end{center}
\end{figure}

Fig.~\ref{fig:noiserate_discharge} shows the noise count rates 
of the PMT for the following setups:

\begin{description}
\item[Setup~1:] reference measurement, with nothing near the PMT window,
\item[Setup~2:] a grounded copper tape attached to
     the PMT window,
\item[Setup~3:] a polycarbonate light-guide, with a surface which was
     deposited by aluminum vapor and coated by silicon oxide, was 
     attached to the PMT window, 
     but electrically isolated from the ground level,
\item[Setup~4:] the light-guide was attached to the PMT window, 
     and grounded, and
\item[Setup~5:] re-measurement of Setup~1, for confirmation.
\end{description}

The PMT was placed in a light-tight box and the noise rate
measured with a threshold discrimination of $-20$~mV, 
corresponding to the single-photoelectron level.
For each setup,
both the cathode- and anode-ground were tested three times.
It was found that setups 2 and 4 with the anode grounded
showed the highest noise rate,
while the noise rate for all cathode-grounded cases was stable.
This shows that the electrical stability of the cathode plane 
is important for the PMT noise rate.

A minor disadvantage of the cathode-ground setup is that signal extraction 
from the PMT anode must be done with an AC coupling, i.e.,
it prevents us from measuring the DC current at the PMT anode.
However, this is not an important consideration
as the threshold level of the pulsed photons is constant
because the DC level of the preamplifier input becomes
stable around the ground level under any circumstance.
Also we monitor a single trigger rate for each PMT by scaler circuit
(in each second with 700 $\mu$s timing gate).
It helps us to know the background light level.

\subsubsection{High-voltage bleeder}
\label{section:pm_bleeder}

\begin{figure}
\begin{center}
\includegraphics[width=70mm]{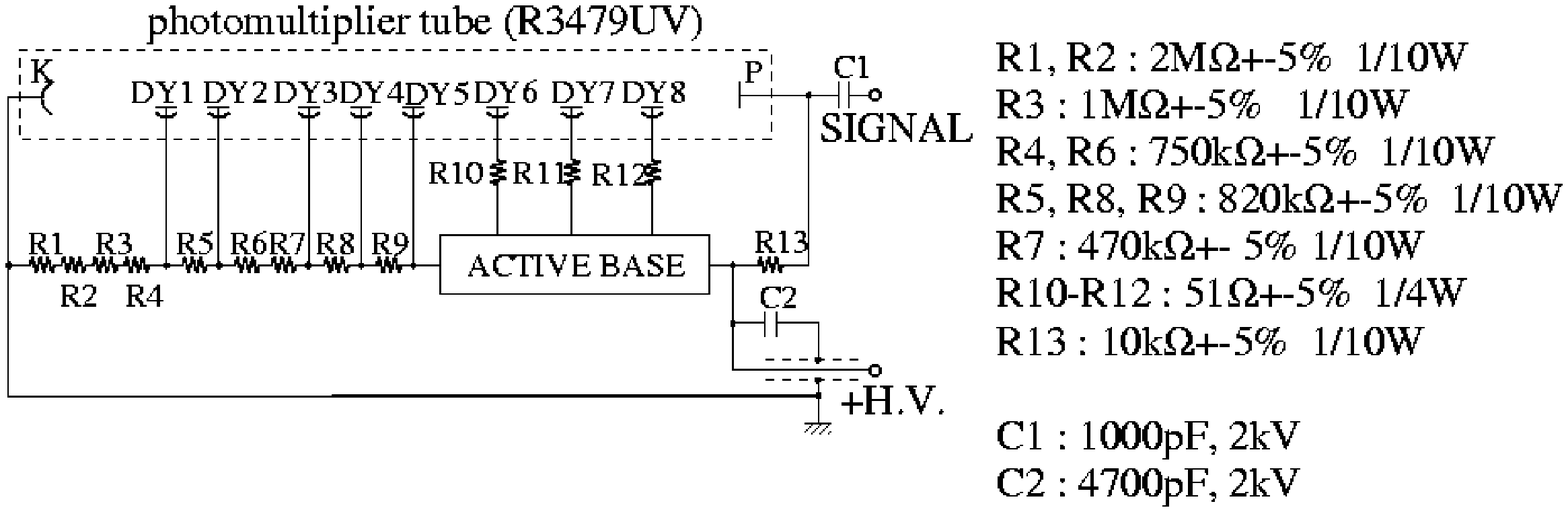}
\vskip 1cm
\caption{Schematic diagram of the high-voltage bleeder.}
\label{fig:diagram_bleeder}
\end{center}
\end{figure}

A schematic diagram of the high-voltage bleeder circuit is shown
in Fig.~\ref{fig:diagram_bleeder}.

An active base, which stabilizes the electric potential difference 
between the dynodes, was introduced at the 6th to 8th locations of the dynodes.
This is to prevent any deterioration of the dynode amplification
when the bleeder suffers from a high constant current flow
due to a huge background of photons, such as the effects of
a bright-star effect (magnitude $<$3).

\begin{figure}
\begin{center}
\begin{minipage}{30mm}
\includegraphics[width=30mm]{activebase-629_mod.epsi}
\end{minipage}
\begin{minipage}{30mm}
\includegraphics[width=30mm]{activebase-627_mod.epsi}
\end{minipage}
\begin{minipage}{30mm}
\includegraphics[width=30mm]{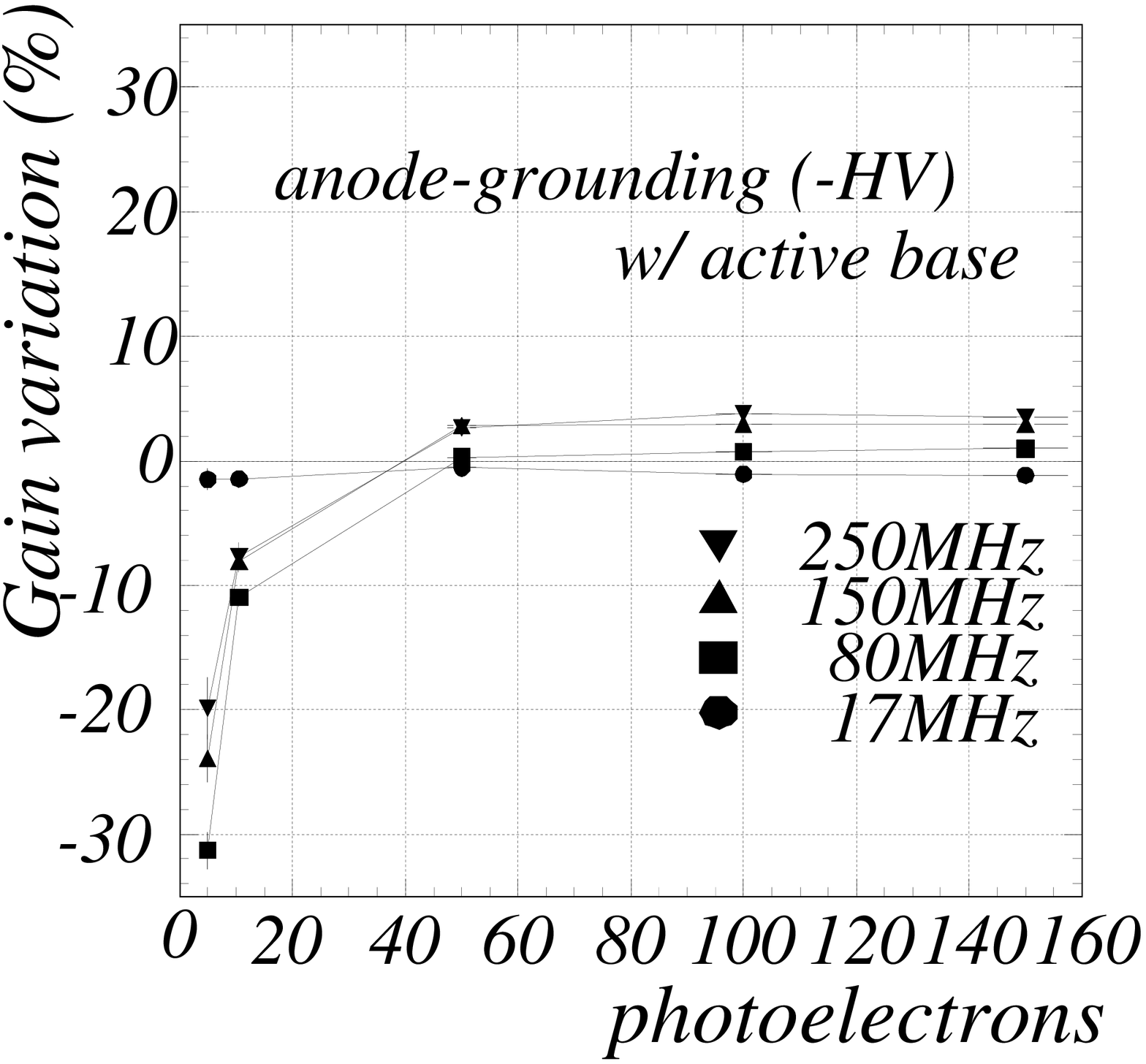}
\end{minipage}
\begin{minipage}{30mm}
\includegraphics[width=30mm]{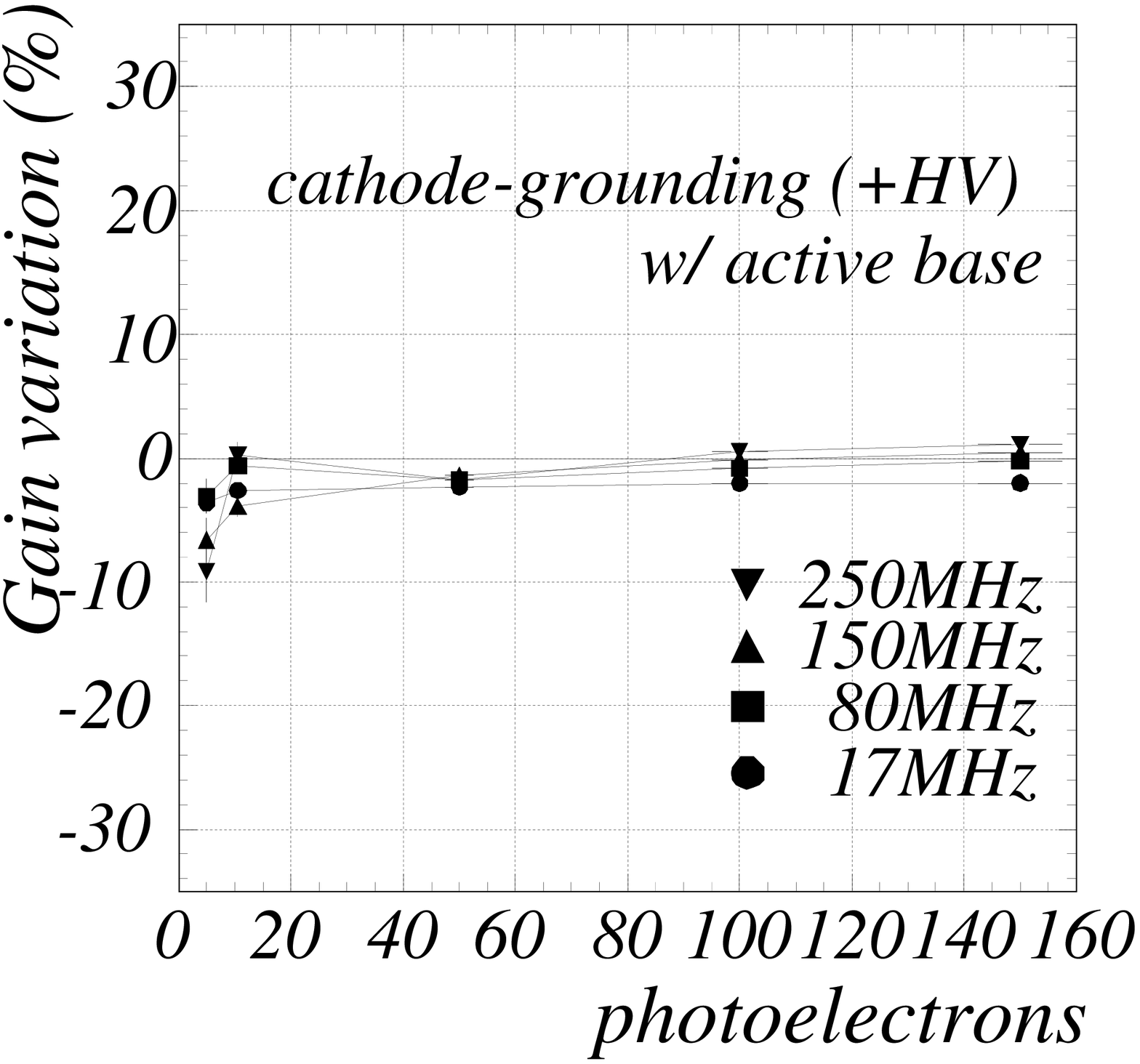}
\end{minipage}
\vskip 1cm
\caption{PMT gains for different amounts of
background photons.
The upper figures show the measurement
for the anode-ground (left) and the cathode-ground (right) setup
for PMTs which were not equipped with active base dynodes.
The lower figures show the cases for the PMTs equipped with an active base.}
\label{fig:HV_polarity_deviation}
\end{center}
\end{figure}

Precise and stable measurements of the pulse height are necessary 
even under a wide variety of
background photon rates.
The feasibility of a photon pulse measurement was investigated 
for several conditions under various background photon rates.
Fig.~\ref{fig:HV_polarity_deviation} shows the difference in 
the pulse height for different levels of background photon fields.

When the incident light level exceeds a certain level, the anode
current began to deviate from the ideal linearity. The detail
of this phenomenon can be found in Ref. \cite{hpkk}.
The typical night-sky background 
is estimated to be $2.6\times10^{-4}$ erg/cm$^2$/s/str~\cite{jelly58}.
Therefore, a single photoelectron rate of $\sim$17~MHz 
for each pixel is expected from  background photons
when the camera is installed at the focal plane of telescope.
We carried out tests with up to 15 times more 
photons than this typical background
and for both cathode- and anode-ground.
From these results, the gain of the PMTs which had active base dynodes,
was found to be stable under an environment of severe background photons.

\subsection{Light-guide}

The camera consists of PMTs with a significant amount of dead 
space between them, amounting to $\sim$65\% of the total surface area.
Light-guides reflect photons which would otherwise be incident upon
the dead space onto the photocathode area of the PMTs,
thus increasing the light-collection efficiency.
Another advantage is that light-guides reduce the background of
photons coming from outside of the mirrors, i.e., at a shallow
angles with respect to the light-guide plane. 
 
After Monte Carlo simulations for various shapes 
of light guides were performed
to evaluate their performance, the optimal 
shape of the light guide was determined.
The most efficient light guide design was based on the Winston cone, 
but with a hexagonal entrance shape.

\subsubsection{Shapes of light guides}

Various types of light guides made by combining the Winston cone, 
paraboloids and flat planes were examined by simulations.
The Winston cone is a non-imaging optical shape used to concentrate 
all photons whose incident angle is less than a certain angle
\cite{win83,win91}. 
The maximum incident angle to the light guide is determined by the
incident light 
from the outer edge of a 10\,m mirror dish, which ranges from 32.2$^\circ$ to 
35.2$^\circ$, depending on the position of the camera surface.
Although the entrance shape of the original Winston cone is round, 
as the PMT's are arranged on a hexagonal grid with a spacing of 24~mm,
a hexagonal entrance aperture was chosen so as to obtain 
less dead space and a better light-collection efficiency.
The ratio of the entrance area and the exit area of the light-guides 
is 2.68 for the case of a gap of 0\,mm between the light-guides 
and 2.46 for a 1\,mm gap.

\subsubsection{Light-collection efficiencies}

Simulations to obtain the light collection efficiency were carried out
based on the following simplified assumptions:
\begin{enumerate}
\item The diameter of the sensitive area of the PMT is 15\,mm. 
\item The distance between the centers of neighboring PMTs is 24 mm. 
\item Reflectances of 100\% and 80\% were trialled for the surface 
      of the light-guide.
\item The reflectance at the surface of the PMT is assumed to be 0\%. 
\item Gaps of 0\,mm and 1\,mm were trialled between neighboring light-guides.
\item Gaps of 0\,mm and 2\,mm were trialled between the light-guide and 
      the PMT surface.
\item Photons are illuminated on the paraboloidal 10\,m mirror 
      and reflected at the mirror surface; they are then injected on a 
      light guide centered at the camera position. 
\item The camera position is assumed to be at the focal plane, 
      8\,m in front of the main reflector.
\end{enumerate}

The simulations showed that, as expected, the collection efficiencies 
varied as the light-guide shape was changed. For example, efficiencies
between 69.4\% to 77.3\% were obtained for 
a reflectance of 80\%, and gaps of 0\,mm.
The best case was obtained for a shape for which every cross
section through the central axis had the shape of Winston cone. 
 
The effects of the gap between the light guides, the gap between the
lightguide and the PMT surface, and the reflectance of the light
guides are given in Table~\ref{table:lgeff}.

\begin{table}
\begin{tabular*}{61mm}{|p{26mm}|r|r|}

\hline
Gap between LGs & 0 mm     & 1 mm \\
\hline
\hline
Reflectance of LG : 100 \%   &  & \\
Gap between LG and PMT : 0 mm & 94.3 \% & 87.0 \% \\
\hline
Reflectance of LG : 80 \%    &  & \\
Gap between LG and PMT : 0 mm & 77.3 \% & 72.2 \% \\
\hline
Reflectance of LG : 80 \%   &  & \\
Gap between LG and PMT : 2 mm & 69.0 \% & 66.4 \% \\
\hline
\end{tabular*}
\vskip 1cm
\caption{Simulated light-collection efficiencies for various cases.}
\label{table:lgeff}
\end{table}

It was found that only a 2mm gap between the lightguide and the PMT
surface reduces the efficiency by nearly 10\%. 

\begin{figure}
\begin{center}
\includegraphics[width=5.3cm]{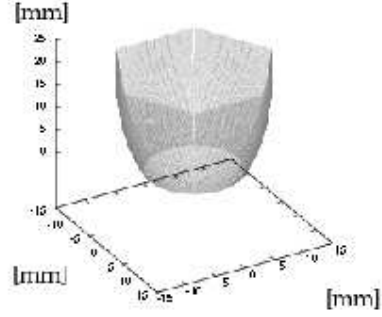} 
\vskip 1cm
\caption{Data points of the inner shape of the new light guide.}
\label{fig:LG_Inner_Shape}
\end{center}
\end{figure}

Based on these results, we 
again 
chose a design for which 
the cross-sections were Winston cones,
the entrance shape was hexagonal, and the exit shape was round. 
A mold was 
turned
with a computer-controlled machine based on the data
shown in Fig.~\ref{fig:LG_Inner_Shape}, and finally polished by hand.

\begin{figure}
\begin{center}
\includegraphics[width=5.3cm]{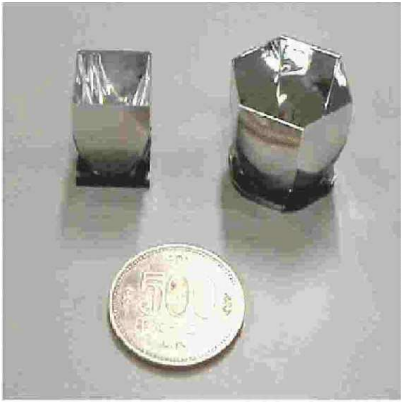} 
\vskip 1cm
\caption{Left : Light guide for the CANGAROO-II telescope.
~~~~Right : New light guide  for the CANGAROO-III telescope.
The coin shown is 26\,mm in diameter. }
\label{fig:LG_picture}
\end{center}
\end{figure}

Polycarbonate plastic was used for the base material.
A picture of the final product is shown in Fig.~\ref{fig:LG_picture}.

The manufacturing error of the size was measured and proved to be less
than 0.1\,mm, however, as there is likely to be some temperature 
dependence, a more conservative value 0.5\,mm was chosen
for the gap between light-guides.
Aluminum vapor was deposited on the inner surface of 
the light guide and the surface was then coated with silicon oxide. 

\begin{figure}
\begin{center}
\includegraphics[width=5.3cm]{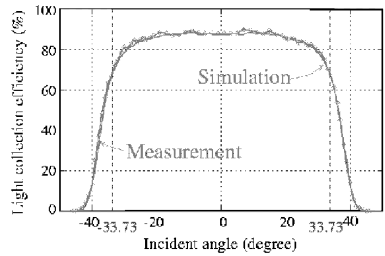} 
\vskip 1cm
\caption{Comparison of light collection efficiency between a simulation and
a measurement of the light-guide.}
\label{fig:LG_efficiency}
\end{center}
\end{figure}

The light-collection efficiency was measured for the manufactured 
light-guide as a function of the incident angle using a blue LED 
with a wavelength of 470\,nm. 
Figure \ref{fig:LG_efficiency} shows the measured efficiency, 
which agrees very well with a simulation 
when the reflectance of the inner surface was assumed to be 89\%.

\subsubsection{Reflectance vs. wavelength}

\begin{figure}
\begin{center}
\includegraphics[width=5.3cm]{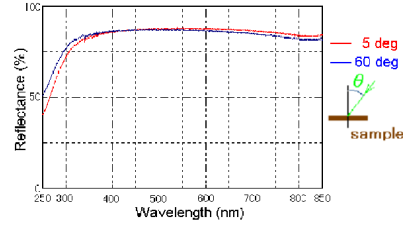}
\vskip 1cm
\caption{Measured reflectance of the new light-guide. }
\label{fig:LG_reflectance}
\end{center}
\end{figure}

Small flat glass samples were put in the same vessel for vapor deposition 
with light-guides.
The reflectance of the samples was measured by a spectrophotometer. 
One of the results is shown in Fig.~\ref{fig:LG_reflectance} as a function
of wavelength. 
The incident angles for this measurement were 5$^\circ$ and 60$^\circ$.
The reflectance value ranges from 75\% to almost 90\% 
above a wavelength of 310\,nm for both incident angles. 

The light-collection efficiency of the light-guide for the camera of the 
second telescope of CANGAROO-III was improved by about 60\%
compared with that of the first (CANGAROO-II) telescope.
The negative high voltages applied to the first telescope camera 
caused discharge problems if the light guides were directly attached
to the PMTs. 
Avoiding this problem requires keeping gaps of about 2\,mm 
between the light guides and the PMTs, which reduces the light-collection
efficiency considerably. 
The second telescope camera employs a positive high voltage to prevent any 
discharge, and so that light guides can be attached directly 
to the surface of the PMTs.

Because the Winston cone shape is used, the light-collection
efficiency of the new light guide drops rapidly at around 35$^\circ$
(Fig.~\ref{fig:LG_efficiency}), which corresponds to 
almost the edge of the 10\,m mirror,  
whereas that for the first telescope changes much more slowly.
Background photons coming from outside of the mirror will also be suppressed 
very much by this property, which will help reducing the trigger rate.
Overall, the light-collection efficiency has been greatly improved
with the new design for the light-guide.

~\

\subsection{High-voltage supply}

A multi-channel and individually controllable high-voltage supply system
is required in order to obtain a uniform pixel gain.
The high-voltage supply system (CAEN SY527) 
controls up to 10 modules of CAEN A932AP,
each of which contains 24 channels of the voltage supply.
The primary voltage of the board can be changed over the range of 0--$2550$~V
and the voltage for the respective channel can be adjusted independently
over the range of $-900$--0~V from the primary voltage,
in 0.2~V intervals.

This high-voltage system can be programmably controlled 
via the VME module, CAEN V288.
A computer-displayed control program 
was constructed using a Tcl/Tk graphical interface, making it 
possible to monitor the status and to control the supplied high voltage.
Also, the monitor program can calculate the position of a bright star image 
on the camera plane using the SAO (Smithsonian Astrophysical
Observatory) catalog 
and accordingly control the gain of the corresponding pixels
on a timescale 
which is much faster than the passage of the 
bright stars around the field of view ($\sim$30~s/pixel).

\subsection{D.C. voltage supply for a preamplifier}

DC voltages ($\pm5$~V) are supplied to the preamplifier circuits.
In order to reduce the voltage drop due to 
the cable loss (247~$\Omega$/km, i.e.\ 6.7~$\Omega$ for 27~m),
the voltages are regulated at the regulator boards 
located inside the camera vessel.
The basic voltages are derived from front-end modules located in
an electronics hut with $\pm12$~V~\cite{kubo01} and 
transmitted through twisted-pair cables.

Each regulator board provides voltages for 16~channels.
A DC noise filter was introduced to reduce the noise due to 
oscillation via the 27~m cables.
A DC fan (size 20$\times$20~mm) is attached to each
regulator chip 
in order to keep temperature variations on 
the regulator board to within 5$^\circ$C.
The noise due to this fan on the output signal was observed to be 
negligibly small.

\subsection{Cables}

The total weight of the cables should be 
less than 70~kg in order to minimize
the deformation of the camera supporting stays, i.e., to keep
the focal plane center within 2~mm.
Therefore, multi-wire cables were selected
instead of coaxial cables 
to transfer signals and to supply high voltages; 
29 twisted-pair cables, 27~meters in length, and 
20 multi-wired cables, 33~meters in length,
were bound to the supporting stays and connected to the camera vessel.
Each twisted pair cable contains 20 pairs. 
The signals from the 16 PMT modules are 
brought together into one twisted-pair cable and 
transferred to the front-end module at the electronics hut~\cite{kubo01}.
The remaining 4 pairs are used to transfer
the DC power for the preamplifier circuits.
The weight of the cables amounts to 130~g per meter.
The rise time of the pulse signal is broadened by the cables
to $\sim$5~ns, but this effect was taken into account in the 
preamplifier circuit, as described in section~\ref{section:preamp}.

High voltages are supplied through multi-wired cables
33~m long and weighing 150~g per meter.
Twenty-four channels can be supplied with one multi-wired cable.
In addition,
two coaxial cables (58C and 174/U), an ethernet cable,
and the AC power supply cables, are also fed into the camera vessel.

\subsection{Calibration light source}

Calibration measurements during an observation are important for 
precise determinations of the properties of gamma-ray showers.
These are carried out by two methods.
The uniformity and absolute value of the gain, and 
the time jitters of the signal pulses,
are calibrated using the following light sources.

\subsubsection{Distant LED light source}
\label{section:distant_LED_source}

The relative-gain uniformity and the timing jitter of the pulse signal are
calibrated using light from an LED; the amount of light and the timing 
are uniformly distributed at the focal plane.
The LED is mounted at the center  
of the telescope dish, 8~meters from the focal plane.
A blue LED (NSPB510S, Nichia),
which emits light of $\sim$470~nm in wavelength, is used.
The light from the LED is oriented about 30$^\circ$ from the optical
axis so that it can be diffused by an optically frosted glass.
The uniformity of the light source on the camera plane 
was measured to be 1.7\% using a CCD (ST-5C, SBIG) in the laboratory.

The advantage of this calibration method is that the data can be taken 
during an observation without changing the experimental setup, 
and so the effect of the background photons due to the night sky 
can be taken into account.
The disadvantage of this method is that we can not see a
single photoelectron peak under the huge night-sky background.

\subsubsection{LED system in the camera vessel }

To monitor the gain of the whole camera system over the longer term, a
new compact monitor system was developed for the camera vessel,
consisting of an LED and a specially patterned screen to diffuse the
light uniformly to every pixel.
The absolute number of photoelectrons by the PMTs can be 
calculated from the widths of the distributions 
of the output charges from the PMTs based on 
Poisson fluctuations.
Using this device and method, the gain of every 
PMT can be obtained 
precisely.  The high uniformity of light over the camera surface
has the advantage of reducing the calibration time. As this device 
is used when the camera lid is closed, 
these operations can be carried out even during the day. 

\begin{figure}
\begin{center}
\includegraphics[width=60mm, angle=0]{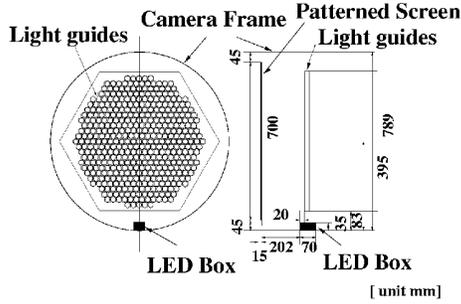}
\vskip 1cm
\caption{Conceptual figure of the camera LED system, which
consists of a light source (LED box) and a patterned screen
installed in the camera vessel.}
\label{fig:Camera_LED_system}
\end{center}
\end{figure}

The calibration unit consists of a patterned 
screen and a LED box containing a blue LED (NSPB520S, Nichia),
a reflector, and optical diffusers 
(shown in Fig.~\ref{fig:Camera_LED_system}).

\begin{figure}
 \begin{center}
  \includegraphics[width=4.5cm]{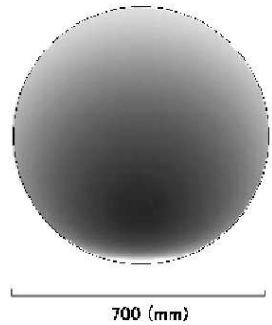}
  \vskip 1cm
  \caption{The patterned screen used to obtain a uniform light intensity
  across the camera from a single LED located at the bottom of the screen.}
  \label{fig:Pattern_screen}
 \end{center}
\end{figure}

The reflector was painted with white VH enamel to diffuse 
the LED light uniformly; the diffusers were 
made of opaque white plastic. 
The patterned screen is placed 22.2~cm in front of 
the light-guides, and is attached to the camera lid. 
The screen is made of polystyrene foam and pattern-printed 
plastic sheets (70~cm in diameter and 1.5~cm thick). 
Black squares 1mm$^2$ in area are printed on the reflector sheet
with a graded density to achieve a uniform distribution of light
over the camera.
The sheet design was based on a Monte-Carlo simulation,
initially assuming that white sheets reflect all incident light and 
that the black toner absorbs it completely,
and taking into account the positions of the light source and the screen,
and the solid angle dependence from the light source.  
The sheets were printed out on a monochrome laser printer. 
After measuring the light intensity of the screen illuminated
by the actual light source, the measured data were fed back into the
simulation. This procedure was repeated iteratively until
good uniformity was obtained. 
The final pattern is shown in Fig.~\ref{fig:Pattern_screen}. 

\begin{figure}
\begin{center}
\includegraphics[width=5.3cm]{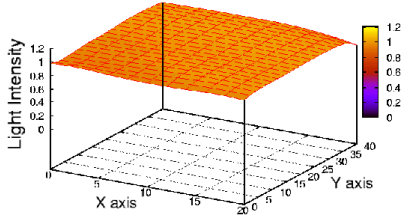}
\vskip 1cm
\caption{Light intensity at the surface of the patterned screen over an
 area of 20~cm $\times$ 40~cm. The average deviation from
uniformity was measured to be 2.6\%.}
\label{fig:Pattern_uniformity}
\end{center}
\end{figure}

The light intensity measured by a single PMT whose position was computer
controlled with an accuracy of about 0.1~mm on an X-Y stage over an area
of 20~cm $\times$ 40~cm, 
as shown in Fig.~\ref{fig:Pattern_uniformity}.

The average deviation from uniformity of the light intensity was 2.6\%.

The average number of photoelectrons ($\mu_{p.e.}$) can be obtained from the
ADC distribution, assuming it obeys
a Poisson distribution, where we assumed that the ADC counts
are proportional to the generated number of photoelectrons. 
The Poisson distribution satisfies the following formula:
 
\begin{equation}
\mu_{p.e.}=\biggl(\frac{{\mu_{ADC}}}{{\sigma_{ADC}}}\biggl)^2 ,
\end{equation}

where $\mu_{ADC}$ and $\sigma_{ADC}$ are the average 
values of the distribution of ADC counts and its standard deviation,
respectively.

\begin{figure}
\begin{center}
\includegraphics[width=5.3cm]{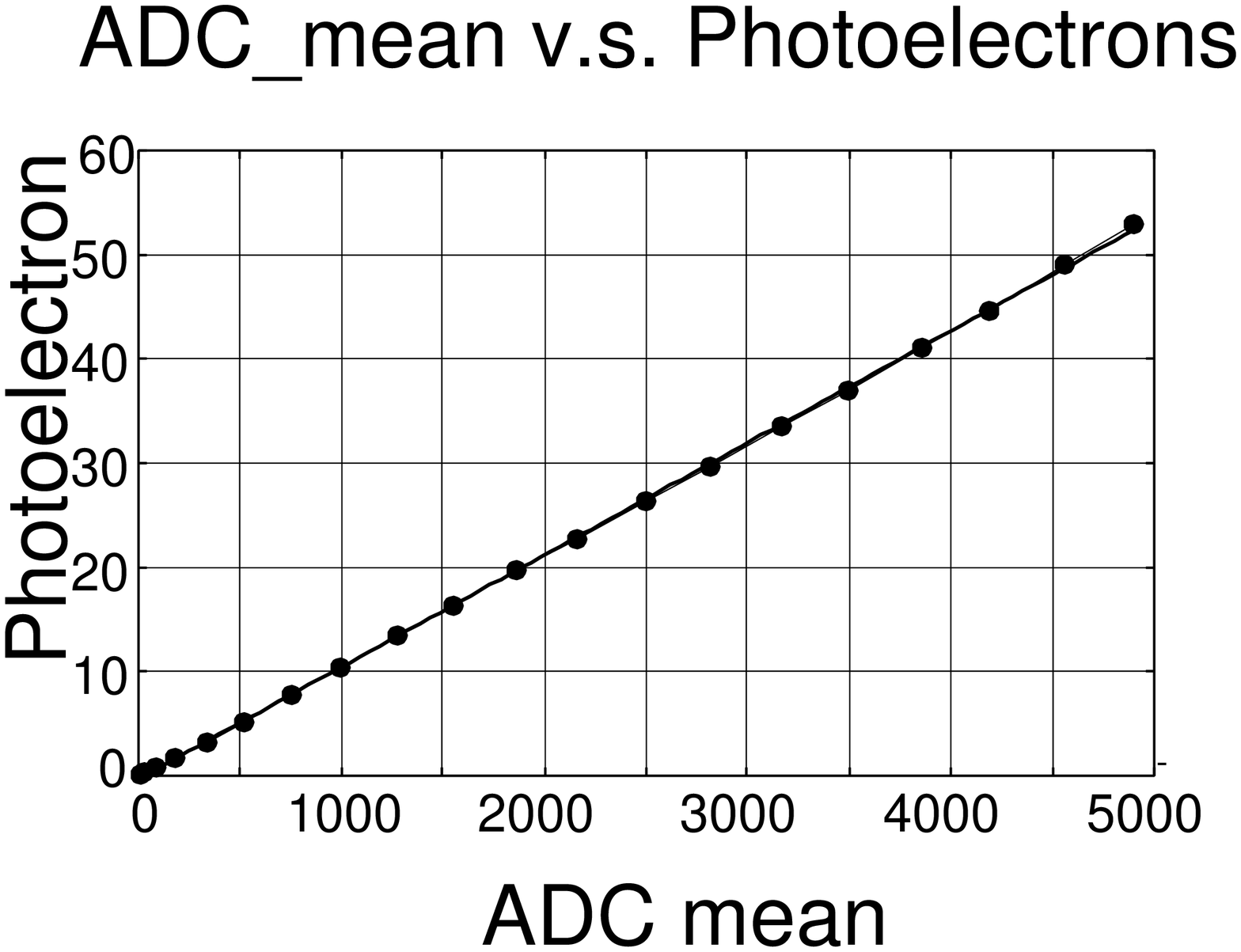} 
\vskip 1cm
\caption{Obtained number of photoelectrons as a function
of the average ADC counts. A straight line fits the data well.}
\label{fig:pe_vs_ADC}
\end{center}
\end{figure}

The linearity of the whole camera system for 427 pixels is measured by this
method. The number of photoelectrons derived using the
above formula is plotted as a
function of the ADC counts in Fig.~\ref{fig:pe_vs_ADC}. 

\begin{figure}
\begin{center}
\includegraphics[width=5.3cm]{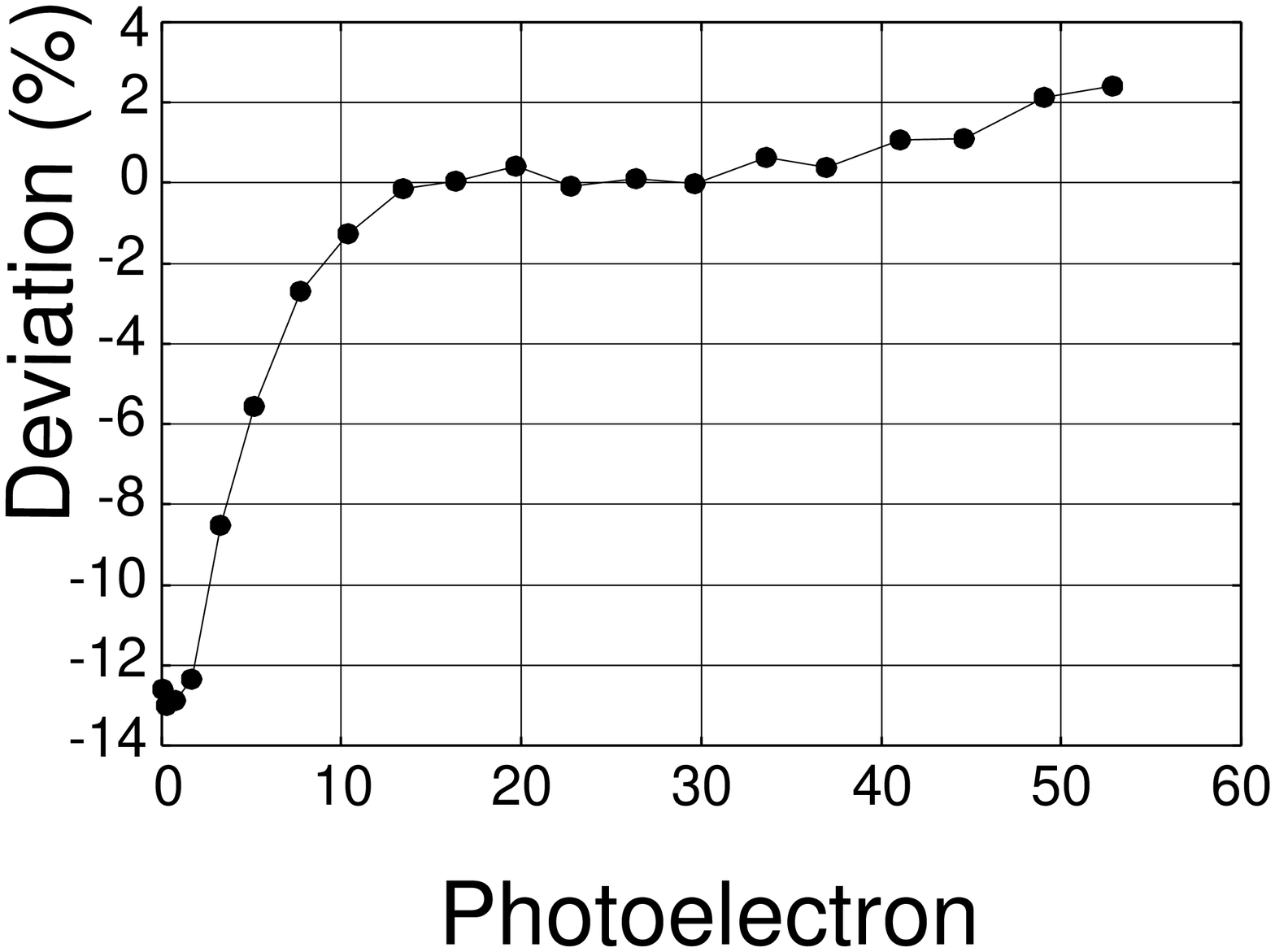} 
\vskip 1cm
\caption{Deviation of data from the fitted straight line.}
\label{fig:Linearity_deviation}
\end{center}
\end{figure}

These points are well fit by a straight line at high input photons.
The deviations of the measured data points from the fitted line is shown in
Fig.~\ref{fig:Linearity_deviation}. 

The deviation is found to be within 2\% from~10 p.e. to 50~p.e..
The larger deviations observed for smaller numbers of
photo-electrons was caused by the finite resolution of the PMTs.

\section{Calibration}

\subsection{Calibration of each PMT}
\label{section:calib_pm_module}

Before the camera was constructed,
the characteristics of all PMT modules were
calibrated individually and the results were stored in a database.
The following properties of the 450 PMT modules, including reserves,
were calibrated:

\begin{itemize}
\item high-voltage dependence of the gain,
\item linearity from 1 to 1000~p.e.,
\item timing resolution, and
\item quantum efficiency.
\end{itemize}

The calibration system was set up to be 
similar to the actual experiment, i.e.,
the output signal was passed through a twisted-pair cable of the 
same length 
in order to take into account the effects due to the cable loss and 
the deterioration of the rise time.

The output signal was split into two, with 
one signal converted by a charge-integrated ADC
with 12~bit resolution and with 1~count corresponding to 0.24~pC,
and the other signal fed into a TDC of 24.4~ps resolution.
Pulses with a width of 20~ns were sent to the LED, and
the amount of light was roughly adjusted by changing the supplied voltage.
Various amounts of light input were able to be distributed 
using combinations of 
neutral density (ND) filters of various transmittance.

\begin{figure}
\begin{center}
\includegraphics[width=50mm]{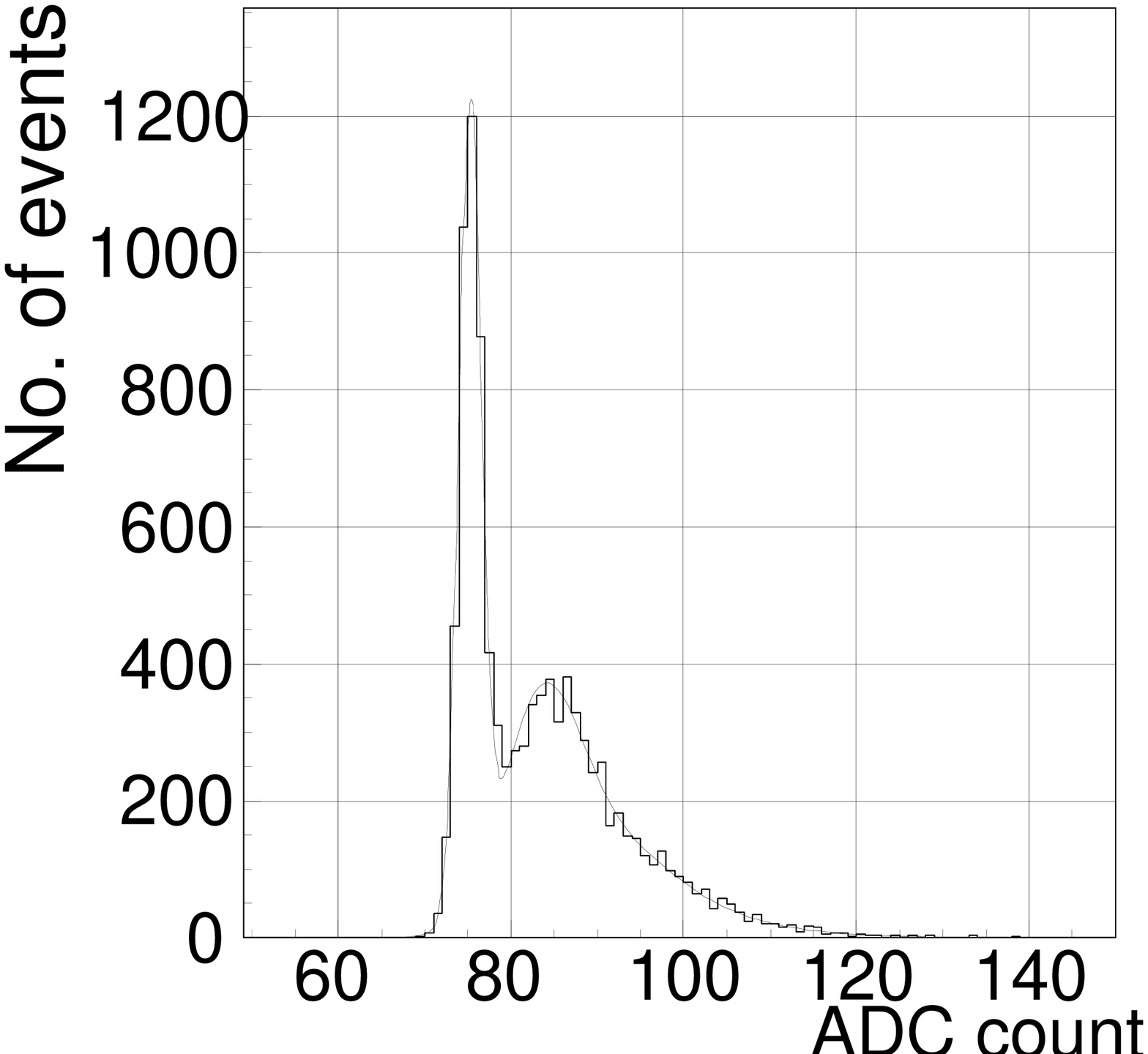}
\vskip 1cm
\caption{Distribution of the single-photoelectron peak
measured by a PMT module.}
\label{fig:histogram_1pe_peak}
\end{center}
\end{figure}

The 450 PMT modules were calibrated by
the following procedure: 
\begin{enumerate}
\item The HV was adjusted to give a gain of
      $1.2\times10^7$ (including the preamplifier gain),
      to measure the single-photon spectrum.
\item The amount of the light was changed to be
      1, 10, 50, 100, 210, 280, 350, 500, 700, and 1000~p.e..
\item The high-voltage dependence of the gain was measured
      with the amount of 10~p.e. input.
\end{enumerate}
Fig.~\ref{fig:histogram_1pe_peak} shows the typical distribution
of a single photoelectron peak.

The peak due to a single photon signal 
can be clearly separated from the background.
Through these calibration processes, five of the 450 PMT modules were 
rejected as they did not meet at least one of the following requirements.

\subsubsection{Gain linearity}

\begin{figure}
\begin{center}
\includegraphics[width=50mm]{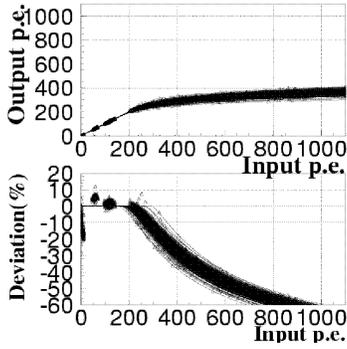}
\vskip 1cm
\caption{Fitted curves (upper) and
deviations from the linear response (lower) for all
PMT modules.}
\label{fig:amplification_linearity}
\end{center}
\end{figure}

Fig.~\ref{fig:amplification_linearity} shows the linearity of 
all the PMT modules for gain (including pre-amplification) of $1.2\times10^7$.

\begin{figure}
\begin{center}
\begin{minipage}{50mm}
\includegraphics[width=50mm]{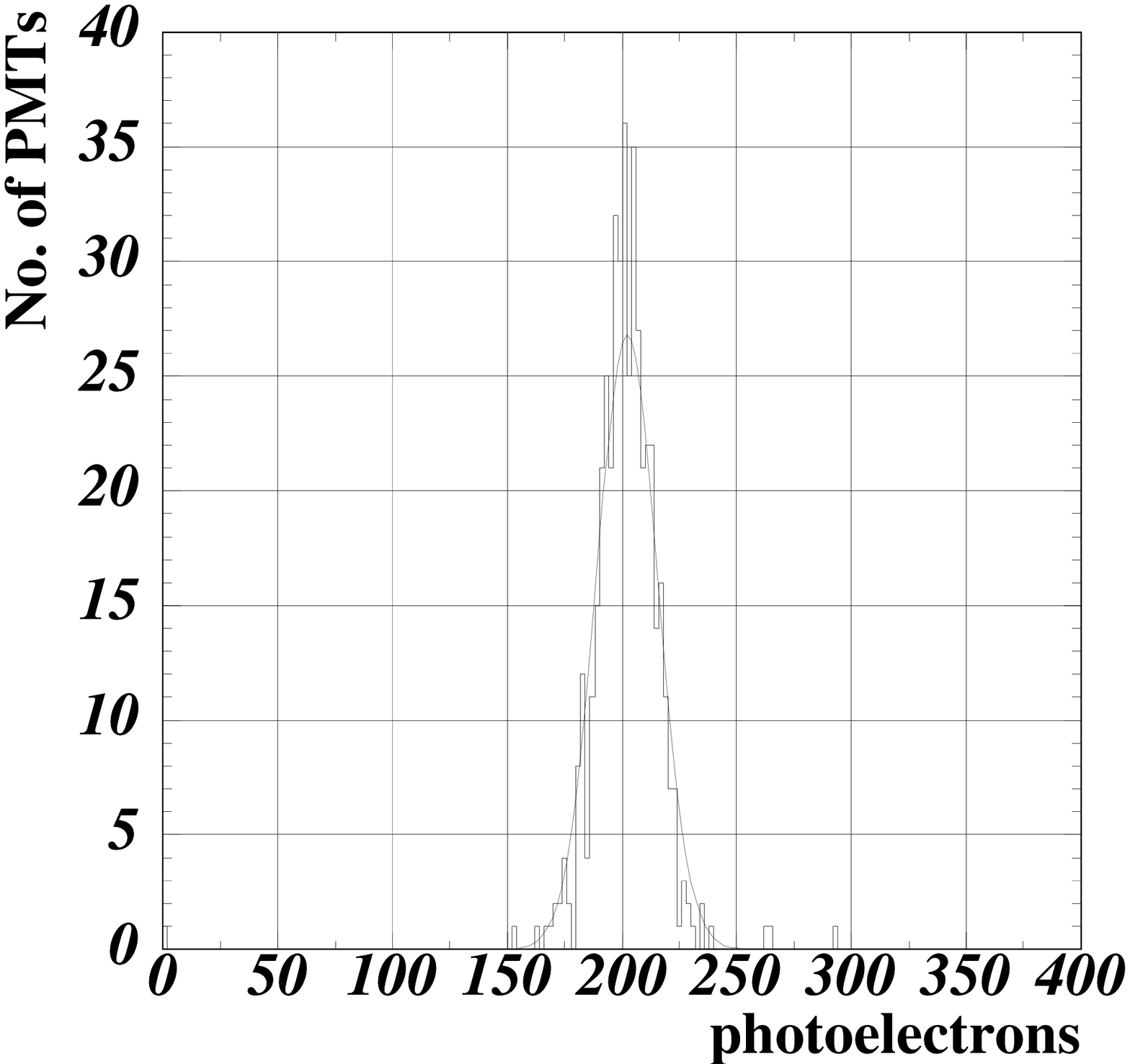}
\end{minipage}
\begin{minipage}{50mm}
\includegraphics[width=50mm]{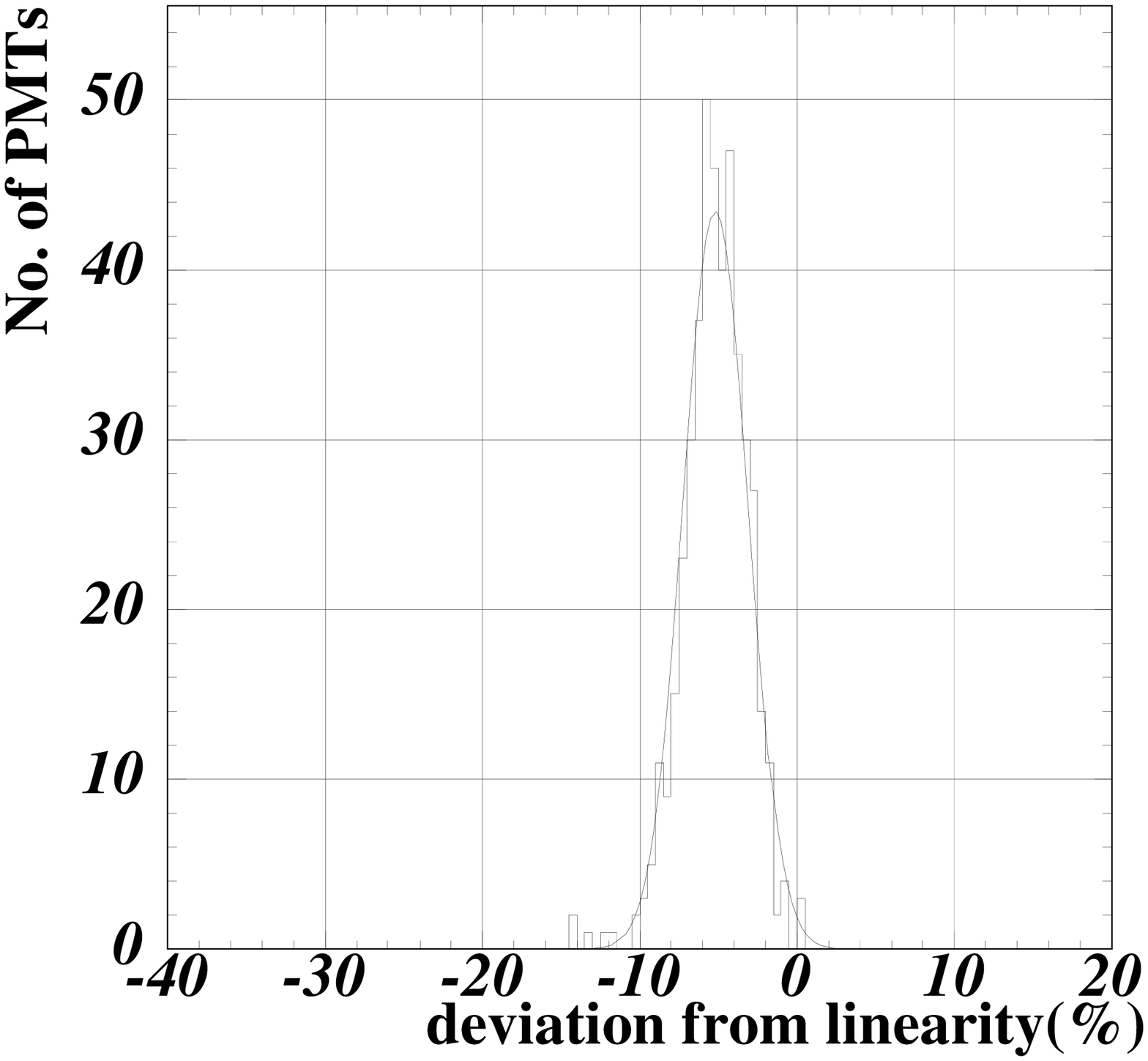}
\end{minipage}
\vskip 1cm
\caption{Distribution of the parameter $a$ (upper) for all PMT modules
and the deviation from the linearity at 250~p.e. of
the input photons (lower).}
\label{fig:saturation}
\end{center}
\end{figure}

Data points were fitted using the following empirical formula:

$$
F(x) = \left\{
            \begin{array}{@{\,}llll}
x & & & \mbox{($x \leq a$)} \\
\frac{((x-a+c)^b-c^b)}{b}c^{(1.0-b)}+a & & & \mbox{($x > a$)}, 
\end{array}
\right. 
$$

\noindent
where $a$ approximately corresponds to 
the turning point of the line. 
The resulting average and standard deviation of $a$ 
were $202.1\pm12.7~(1\sigma)$~p.e., 
as shown in Fig.~\ref{fig:saturation} (upper).

The deviation from a linear line at 250~p.e. of the input light was 
estimated to be $-5.1\pm2.0~(1\sigma)$~\% 
from Fig.~\ref{fig:saturation} (lower).
PMT modules with a deviation worse than $-$20\% were rejected.

\subsubsection{High-voltage dependence of gain}

\begin{figure}
\begin{center}
\begin{minipage}{50mm}
\hskip 0.5cm\includegraphics[width=50mm]{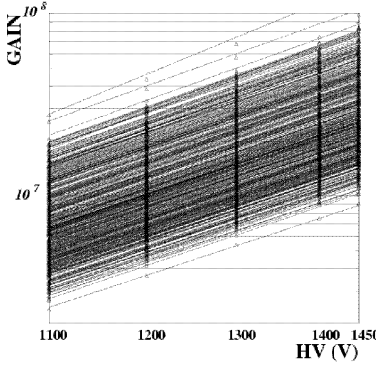}
\end{minipage}
\vskip 1cm
\begin{minipage}{50mm}
\includegraphics[width=50mm]{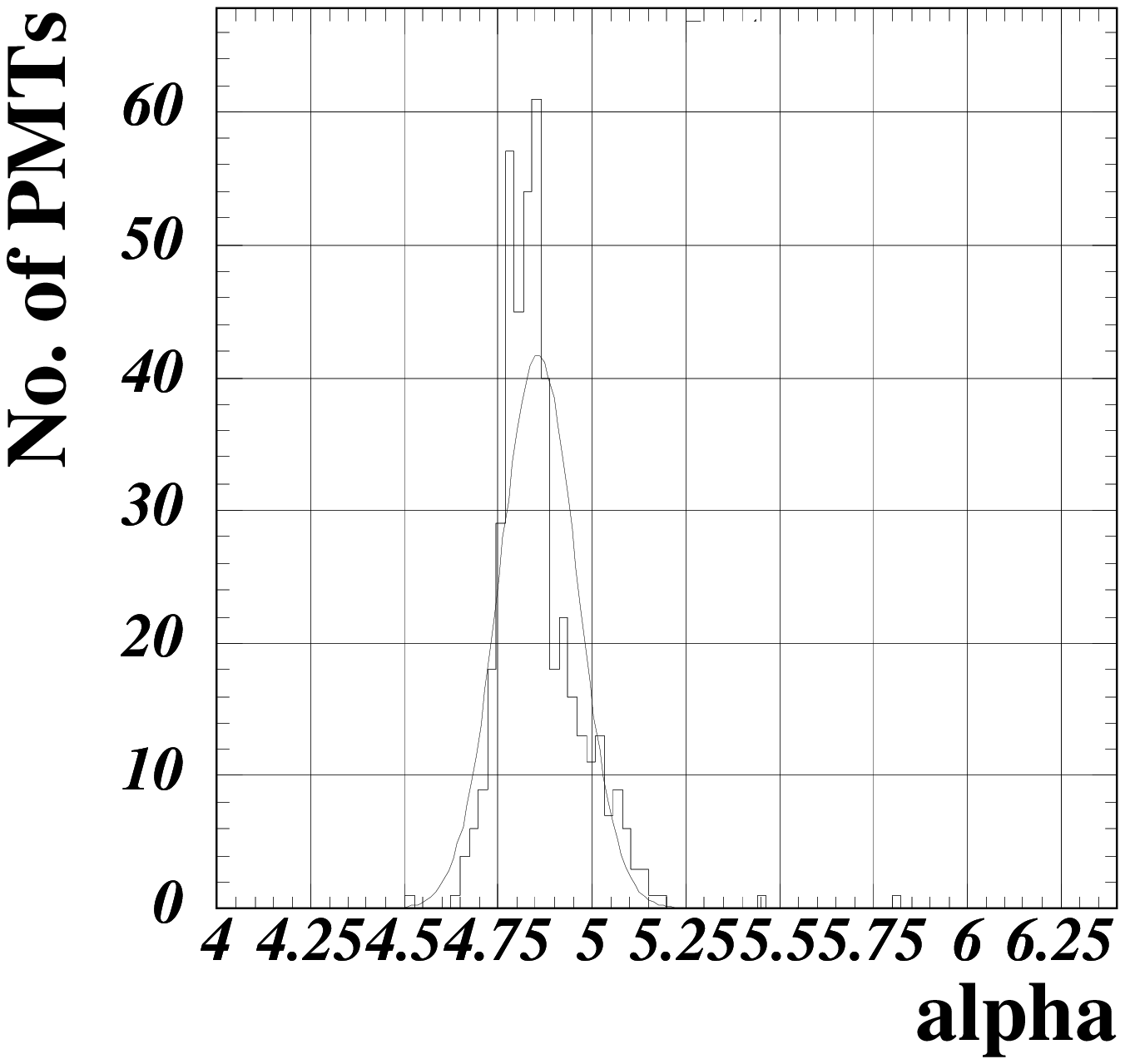}
\end{minipage}
\vskip 1cm
\caption{The upper figure shows the gain versus high voltage
for all PMT modules.
The data points were fitted with the function $K \cdot V^\alpha$.
The lower figure shows the distribution of parameter $\alpha$,
obtained from the fitted line.}
\label{fig:hv_kalpha}
\end{center}
\end{figure}

Fig.~\ref{fig:hv_kalpha} (upper) shows the high-voltage dependence of 
the gain. 

The gain was measured at HVs of
1100, 1200, 1300, 1400 and 1450~V,
and fitted to the following formula:

$$
(\mbox{Gain}) = K \cdot V^\alpha,
$$

\noindent
where $V$ is the supplied high voltage and $K$ and $\alpha$ are
free parameters.
The parameter $\alpha$ represents the gain sensitivity for the high voltage,
and the average value for all modules was
$4.9\pm0.1$, as shown in Fig.~\ref{fig:hv_kalpha} (lower).
PMT modules with $4.5<\alpha<5.25$ were selected in
order to keep the operating HV within the controllable
range.
Using these fit parameters, it is possible for
observations to be conducted with the appropriate gain
achieved by changing the HV according to the values in the database.

\subsubsection{Timing resolution}

The timing resolution of each PMT module was estimated 
after a correction for the time-walk effect.
The discrimination timing can be solved analytically 
when the shape of the signal pulse is approximated by
a Gaussian distribution, $\exp(-at^2)$ and
the timing correction can be applied according to the following formula:

$$
\mbox{T}_{\mbox{start}} = \sqrt{b\log(\mbox{ADC})-c}.
$$

About 10,000 events were obtained for each measurement, and
the parameters $b$ and $c$ were obtained by the fitting procedure.
For a few PMTs, full measurements were carried out, and 
each PMT was confirmed to have less than a 1~ns time resolution.
When the input signals were greater than 20~p.e.,
this correction became unnecessary.
Therefore we undertook the rest of the time-resolution measurements
with a 20~p.e. input (and without the time-walk correction) 
for the remainder of the PMT modules.

\begin{figure}
\begin{center}
\includegraphics[width=50mm]{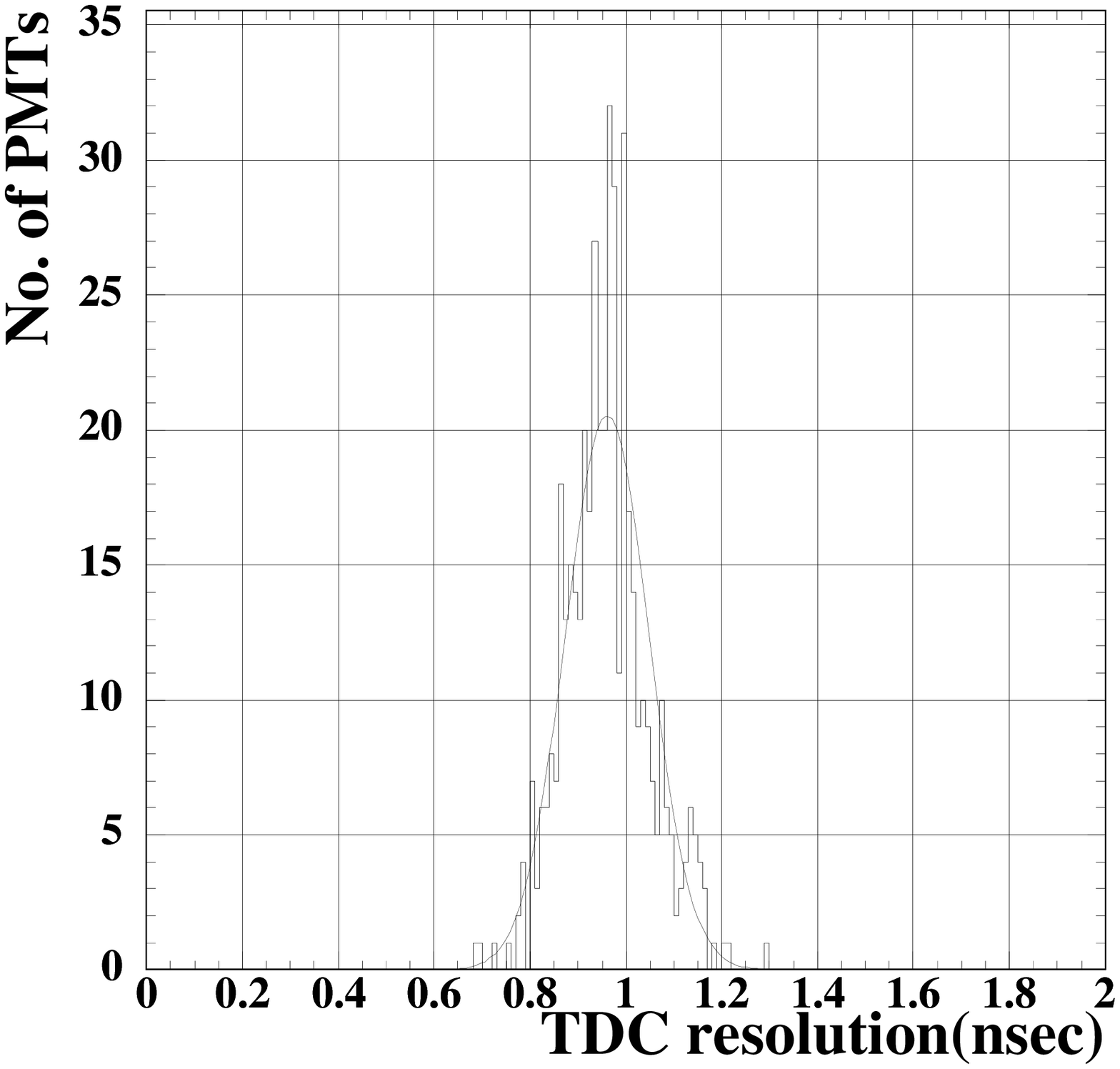}
\vskip 1cm
\caption{Timing resolution for 450 PMT modules when 20~p.e. were
irradiated.}
\label{fig:timing_resolution}
\end{center}
\end{figure}

The resulting timing resolution for all PMT modules is shown in 
Fig.~\ref{fig:timing_resolution}.

The average value was $0.96\pm0.09$ ns.
It should be noted that this resolution was obtained under
the effect of a deterioration of the rise-time due losses in the
27-meter cable.

\subsubsection{Estimation of the quantum efficiency}
\label{section:quamtum_efficiency}

\begin{figure}
\begin{center}
\includegraphics[width=50mm]{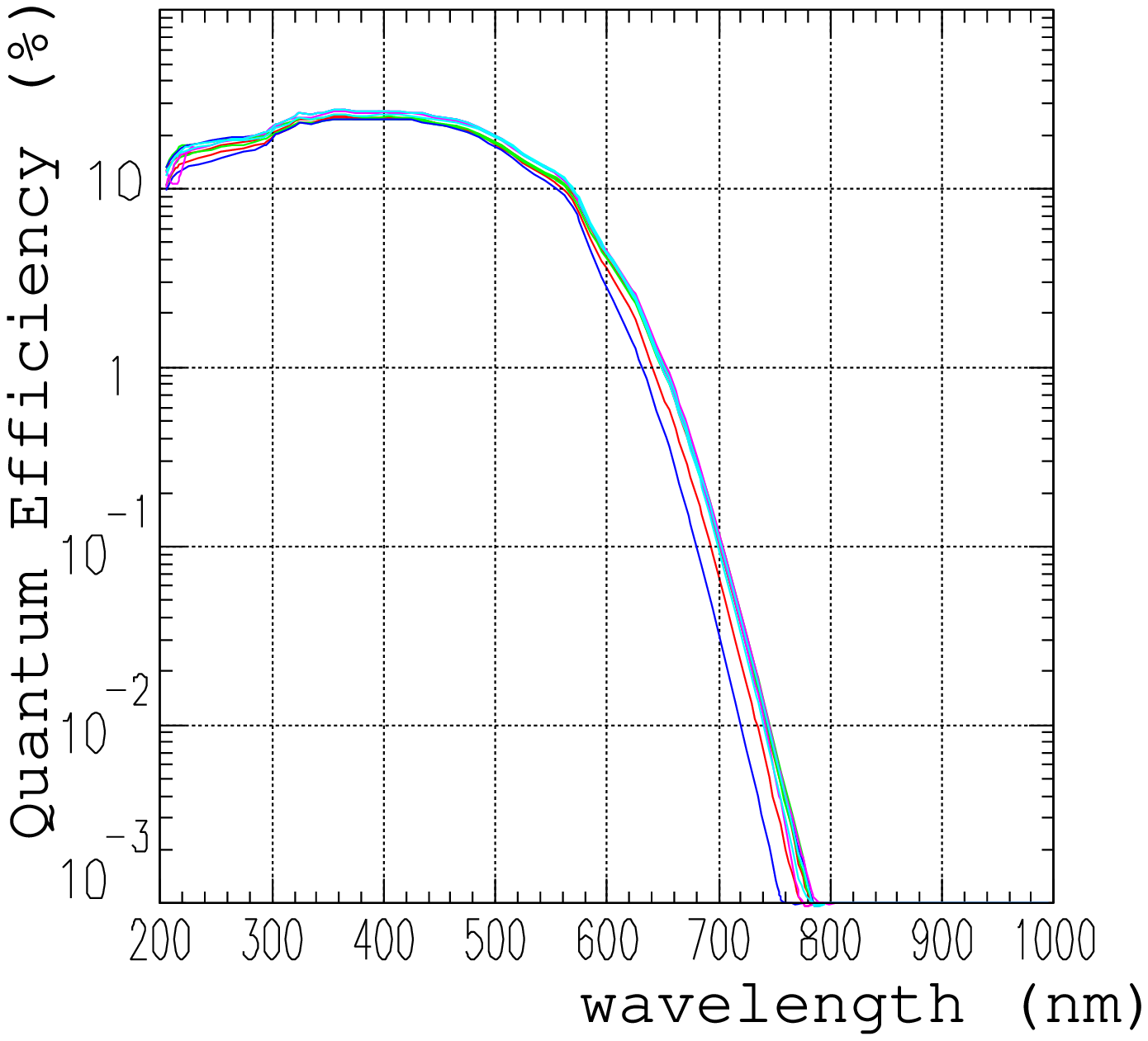}
\vskip 1cm
\caption{Quantum efficiency as a function of the wavelength
measured for 10 PMTs.}
\label{fig:quantum_efficiency}
\end{center}
\end{figure}

The quantum efficiencies were measured for 10 of the 450 PMTs 
as a function of the wavelength, 
as shown in Fig.~\ref{fig:quantum_efficiency}.

\begin{figure}
\begin{center}
\includegraphics[width=50mm]{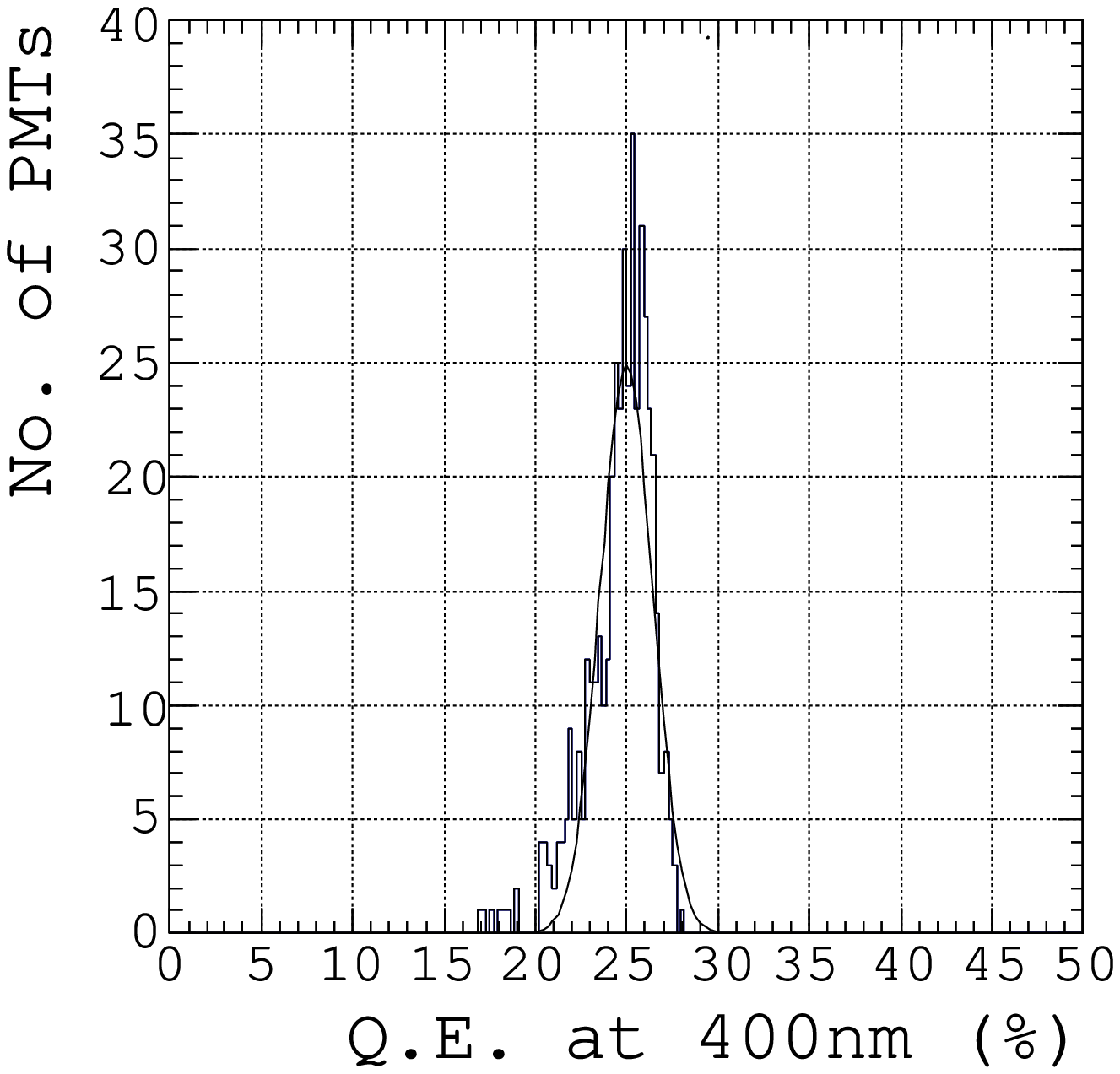}
\vskip 1cm
\caption{Estimated quantum efficiencies using $Skb$ values.}
\label{fig:quantum_efficiency_450}
\end{center}
\end{figure}

The efficiency curves 
were very similar for all 10 PMTs.
Since it is difficult and expensive to 
measure the quantum efficiency for many PMT modules,
the quantum efficiencies were estimated for all modules from the  
$Skb$ parameter, which gives the sensitivity of the cathode
for blue light.
$Skb$ was measured with light from a Tungsten lamp filtered
with an optical filter to a narrow wavelength band around 400~nm.
Its unit is [A/lm].
It is easier to use $Skb$ than to measure the quantum efficiency.
We confirmed that there is a good correlation between 
the quantum efficiency at 400~nm 
and $Skb$ for the 10 PMTs for which full measurements were made,
and then estimated the quantum efficiency 
based on the measured $Skb$ and the relation derived from 10 PMT sample,
which is shown in Fig.~\ref{fig:quantum_efficiency_450}.

The average of the quantum efficiency was estimated to be
$25.0\pm1.4$\% (a variation of 7.3\%).

\subsection{Performance of the whole camera system}

The total performance of the camera was measured in the laboratory
before it was moved to the experimental site.
The blue LED calibration source 
described in section~\ref{section:distant_LED_source}, 
was placed 8~m 
in front of the camera.
The camera vessel was fixed at two positions of the mounting frame, 
and was 
tilted
over the range of $0^\circ\sim50^\circ$.

High voltages were supplied so as to obtain gains after pre-amplification
of $1.2\times 10^7$
according to the database derived from the calibrations 
described in section~\ref{section:calib_pm_module}.
These measurements were made to check:
\begin{itemize}
\item the uniformity measurement of the gain,
\item the incident-angle dependence of the acceptance efficiency, and
\item cross-talk effects.
\end{itemize}

\subsubsection{Uniformity of gain}

The uniformity of the gain for all pixels was measured with the diffused 
LED light source 
described in section~\ref{section:distant_LED_source}.
The amount of incident light to each pixel area
was adjusted to be about 50~p.e.
The uniformity of the light is described in 
section~\ref{section:distant_LED_source}.

\begin{figure}
\begin{center}
\begin{minipage}{50mm}
\includegraphics[width=50mm]{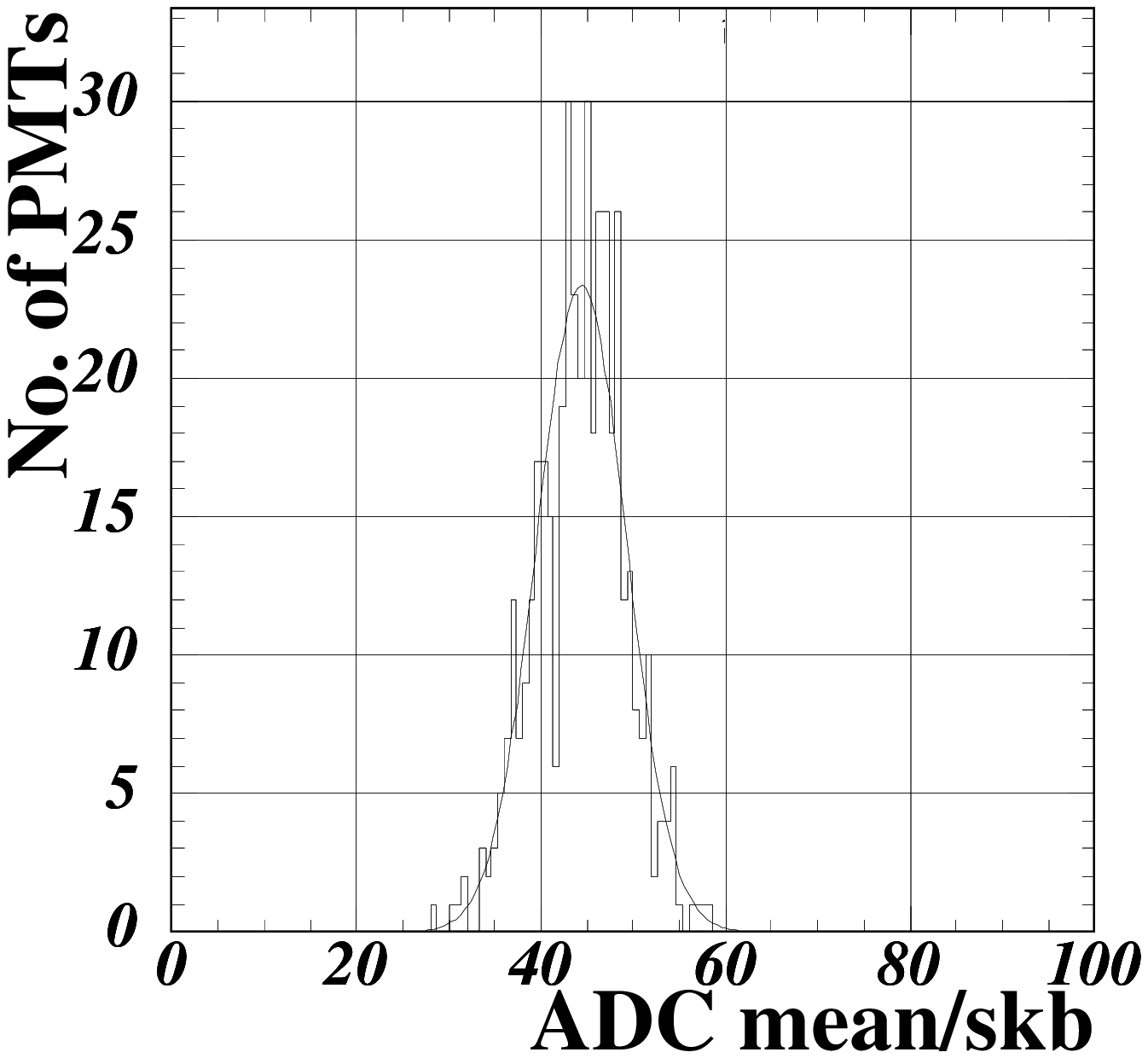}
\end{minipage}
\vskip 1cm
\caption{Gains for all PMT modules in the camera.}
\label{fig:total_gain_dispersion}
\end{center}
\end{figure}

Fig.~\ref{fig:total_gain_dispersion} shows 
the uniformity of the gain for all PMT modules.

We should note that the measured gain by this calibration
should be slightly different from that measured by a single~p.e. calibration
because the dependence of the quantum efficiency for the respective PMTs 
cannot be estimated by a single~p.e. calibration.
Therefore, the measured ADC value should be divided by the $Skb$ value,
which can be considered to correspond to the quantum efficiency.
The average ADC/$Skb$ over all PMT modules was $44.4\pm4.8$
[counts$\cdot$lm/A], 
a deviation of 11\%.

\subsubsection{Incident-angle dependence of the photon acceptance }

\begin{figure}
\begin{center}
\includegraphics[width=50mm]{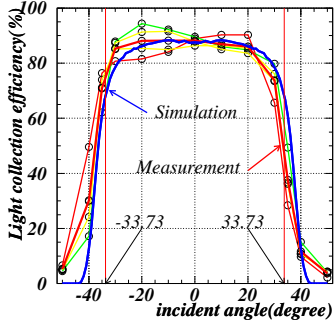}
\vskip 1cm
\caption{Measured and expected
collection efficiency as a function of the incident angle of light.
The results with different rotation angles around the PMT axis
are shown by four lines;
the thick blue line shows their average.}
\label{fig:angle-1}
\end{center}
\end{figure}

At first, the angle dependence was measured for a single PMT module.
The dependence was measured for several incident angles between 
$-50^\circ$ and $+50^\circ$.
For each measurement, the PMT module was rotated by 
90$^\circ$, 180$^\circ$ and  270$^\circ$
around the axis of the PMT module,
in order to average the position dependence of 
the quantum efficiency on the PMT cathode plane.
The efficiency of the light acceptance was defined from
the difference of ADC counts measured with and without light guides
after a correction of the difference of the front/back area of the light
guides, 
which is calculated as follows: 

\[
\mbox{Efficiency} = 
{\frac{\mbox{ADC}_{1}}{\mbox{ADC}_{2}}}/
{\frac{S_1}{S_2}},
\]

\noindent 
where $\mbox{ADC}_{1}$ ($\mbox{ADC}_{2}$) is
the ADC counts with (without) the light-guide,
and $S_1$ and $S_2$ are the area size of the front and back planes of 
the light guides ($S_1$/$S_2$=2.57), respectively.
Fig.~\ref{fig:angle-1} shows the light-correction efficiency estimated from
the measurement. The simulations and data agree well.

\begin{figure}
\begin{center}
\includegraphics[width=50mm]{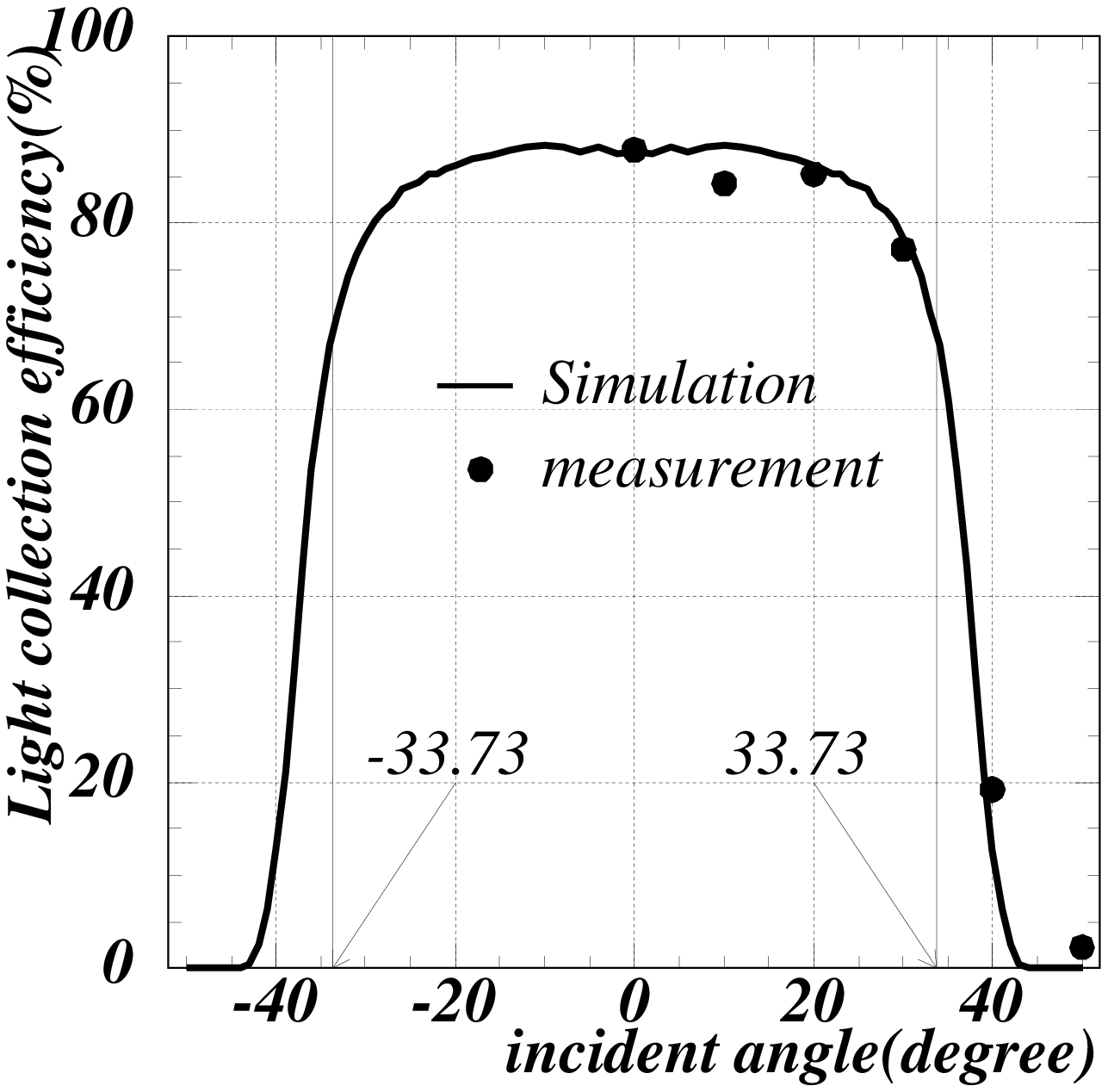}
\vskip 1cm
\caption{Incident-angle dependence of the light collection efficiency
for the whole camera system.
The data points show the measured data and the line shows the expectation
from simulations.
The absolute efficiency of the measurement was adjusted to
that of the simulation at $0^\circ$ incident angle.}
\label{fig:angle-2}
\end{center}
\end{figure}

Fig.~\ref{fig:angle-2} shows the dependence on incident angle of 
the light-collection efficiency with the whole camera system.

The relative acceptance was estimated with ADC count 
summed for all PMTs, 
and the absolute efficiency was adjusted to 
that of the simulation at an incident angle of 0$^\circ$.
According to this measurement,
the angle dependence of the light collection efficiency agrees well
with expectations.

\subsubsection{Cross-talk effect}

\begin{figure}
\begin{center}
\begin{minipage}{50mm}
\includegraphics[width=50mm]{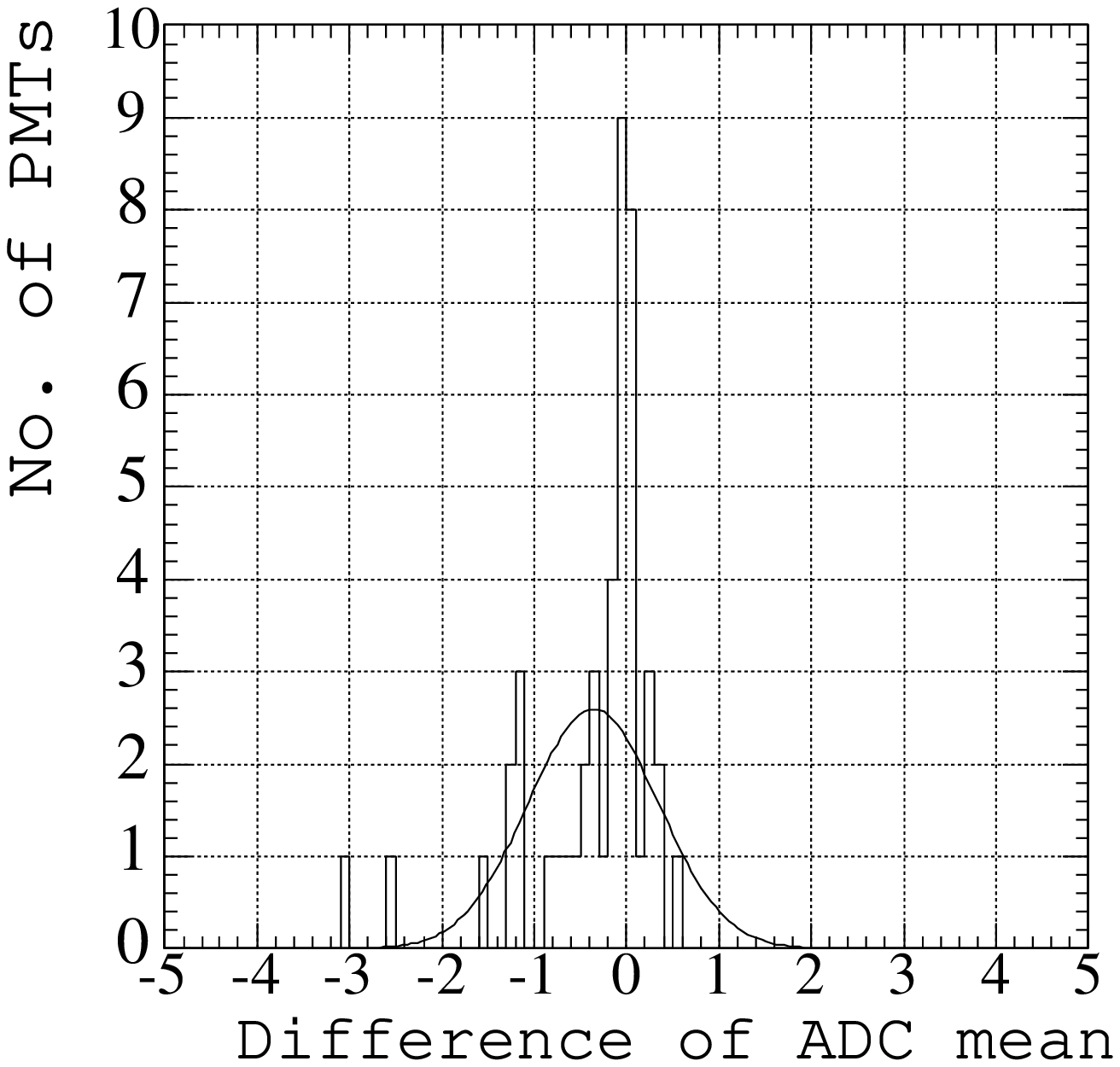}
\end{minipage}
\vskip 1cm
\caption{Deviation of the ADC count for non-illuminated PMT modules.}
\label{fig:cross_talk}
\end{center}
\end{figure}

The cross-talk effect among the neighboring pixels
was investigated by illuminating one PMT module
located at the center of the camera
with an LED (at about 100~p.e. level).
The effect of the cross-talk was investigated while considering 
the variation of the signals for 47 non-illuminated PMT modules
located within a distance of three pixels 
from the illuminated PMT module.
Fig.~\ref{fig:cross_talk} shows the difference of the ADC count for 
a non-illuminated PMT module.

The averaged difference for these PMT modules resulted in $-0.37$~count,
which corresponds to a cross-talk of less than 0.4\%, a negligibly small
value.

Two causes of cross-talk can be considered: 
the location dependence of the PMT module and cross-talk in the signal
cables and connectors.
For the former, no apparent  dependence on  location
was observed, with the cross-talk being less than 0.08\%.
For the latter cause a deviation of 0.1\% was observed,
however, this can also be considered to be negligibly small.

%
%
%
%

\section{Conclusion}

The performance of the Cherenkov imaging camera 
for the second CANGAROO-III telescope was improved 
over that of the first (CANGAROO-II) telescope
with respect to 
the uniformity of gain, timing resolution
and the light-collection efficiency.
The 427 camera PMTs and a number of spares have been
carefully calibrated and the results stored in a database
so that the camera performance can be optimized.
The total weight of the camera was measured to be 110~kg.
This camera performance matches the expectations from
simulations, and is suitable for observations
with the next generation of gamma-ray telescopes.

\section*{Acknowledgments}

We are very grateful to Hamamatsu Photonics K.K.
This work was supported by a Grant-in-Aid for
Scientific Research of Ministry of Education,
Culture, Science, Sports and technology of Japan,
Australian Research Council, and Sasagawa
Scientific Research Grant from the Japan Science Society.
K.T.\ and K.O.\ acknowledge the receipt of JSPS Research Fellowships.
Members of Konan University are grateful for partial support from the 
Promotion and Mutual Aid Corporation for Private Schools of Japan, and
also for assistance with the reflectance measurements for the 
light-guides by Kamoshita Co.\ Ltd.


\begin{thebibliography}{}

\bibitem{hara93}
T. Hara, et al., Nucl. Instr. and Meth. A 332 (1993) 300.

\bibitem{kifune95}
T. Kifune, et al., Astrophys. J. 438 (1995) L91.

\bibitem{tanimori98a}
T. Tanimori, et al., Astrophys. J. 492 (1998) L33.

\bibitem{okumura02}
K. Okumura, et al., 2002, in preparation.

\bibitem{tanimori98b}
T. Tanimori, et al., Astrophys. J. 497 (1998) L25.

\bibitem{muraishi00}
H. Muraishi, et al., Aston. Astrophys. 354 (2000) L57.

\bibitem{enomoto02b}
R. Enomoto, et al., Nature, 416 (2002) 823.

\bibitem{morim01}
M. Mori, et al., in Proc.\ 27th ICRC (Hamburg, Germany), 
5 (2001) 2831.

\bibitem{kubo01}
H. Kubo, et al., in Proc.\ 27th ICRC (Hamburg, Germany), 
5 (2001) 2900.

\bibitem{kajino01}
F. Kajino, et al., in Proc.\ 27th ICRC (Hamburg, Germany), 
5 (2001) 2909.

\bibitem{kawachi01}
A. Kawachi, et al.,  Astropart. Phys., 14 (2001) 261.

\bibitem{hegra97}
A. Daum, et al., Astropart.\ Phys. 8 (1997) 1.

\bibitem{enomoto02a}
R. Enomoto, et al., Astropart.\ Phys. 16 (2002) 235.

\bibitem{akerlof91}
C.W. Akerlof, et al., Astrophys. J. 377 (1991) L97.

\bibitem{hoffman97}
Hoffman, W. 1997, in Proc.\ Towards a Major Atmospheric Cherenkov
Detector V, (Kruger Park, South Africa) p.284.

\bibitem{EGRET}
R.C. Hartman, et al., Astrophys. J. Supp. 123 (1999) 79.

\bibitem{hikishov62}
A.I. Nikishov,  Sov.\ Phys.\ JETP 14 (1962) 393.

\bibitem{gould67}
R.J. Gould, \& G. Schr\`{e}der, Phys.~Rev. 155 (1967) 1408.

\bibitem{stecker92}
F.W. Stecker, O.C. De Jager, M.H. Salamon, Astrophys. J. 390 (1992) L49.

\bibitem{hayashida98}
H. Hayashida, et al., Astrophys. J. 504 (1998) L71.

\bibitem{lampeitl99}
H. Lampeitl, et al., in Proc.  
Towards a major atmospheric Cherenkov detector (Snowbird 1999) 328-332.

\bibitem{grindlay75}
J.E. Grindlay, et al., Astrophys. J. 201 (1975) 82.

\bibitem{hillas82}
A.M. Hillas, J.~Phys. G: Nucl. Phys. 8 (1982) 1475.

\bibitem{reynolds93}
P.T. Reynolds, et al., Astrophys. J. 404 (1993) 206.

\bibitem{khan98}
M.H.R. Khan, Nucl. Instr. and Meth. A 413 (1998) 201.

\bibitem{tsukada91}
K. Tsukada, et al., Nucl. Instr. and Meth. A 300 (1991) 575.

\bibitem{morim99}
M. Mori, et al., in Proc.\ 26th ICRC (Salt Lake City, USA), 
5 (1999) 287.

\bibitem{hpkk}
Hamamatus Photonics K.K., ``Photomultiplier Tubes, Basics
and Applications (second edition)'' p159-160 and Fig. 7-4, 
Hamamatsu Photonics K.K.

\bibitem{jelly58}
J.V. Jelly, 1958, ``Cherenkov Radiation and its Applications'', Pergamon Press.

\bibitem{win83}
R. Winston, \& J. O'Gallagher, Int. Journal of Ambient Energy 4 (1983) 171.

\bibitem{win91}
R. Winston, Scientific American 264 no.3 (1991) 52.


\end{thebibliography}
\end{document}